\documentclass[final]{siamltex}

\usepackage[latin1]{inputenc} 
\usepackage[T1]{fontenc} 
\usepackage{graphicx}
\usepackage{multicol}

\usepackage{subfig}

\def\aalm{{$\bm{a}_{\ell m}$}}
\def\blm{{$\bm{\beta}_{\ell m}$}}
\def\miu{{$\bm{\mu}_{m}$}}

\def\s2hat{{\sc s$^2$hat}}
\def\alm2map{{\tt alm2map}}

\def\deltam{{$\bm{\Delta}_{m}\ $}}

\def\l#1{\left#1}
\def\r#1{\right#1}
\def\simlt{\lower.5ex\hbox{$\; \buildrel < \over \sim \;$}}
\def\simgt{\lower.5ex\hbox{$\; \buildrel > \over \sim \;$}}

\usepackage{bm}
\usepackage{algorithmic}
\usepackage[ruled]{algorithm}

\title{Spherical harmonic transform with GPUs}

\author{{Ioan Ovidiu HUPCA}
\thanks{INRIA Saclay-Ile de France, Bat 490, Universit\'e Paris-Sud 11, France ({\tt ioanovidiu.hupca@lri.fr}). } 
\and{Joel FALCOU}
\thanks{Laboratoire de Recherche en Informatique, Bat 490, Universit\'e Paris-Sud 11, France ({\tt joel.falcou@lri.fr }). } 
\and{Laura GRIGORI}
\thanks{INRIA Saclay-Ile de France, Bat 490, Universit\'e Paris-Sud 11, France ({\tt laura.grigori@inria.fr}). }
\and{Radek STOMPOR}
\thanks{CNRS, Laboratoire Astroparticule et Cosmologie, Universit\'e Paris Diderot, France ({\tt radek@apc.univ-paris7.fr }).   } 
}

\begin{document}

\maketitle

\begin{center}
INRIA Technical Report 7409
\end{center}

\begin{abstract} 

We describe an algorithm for computing an inverse spherical harmonic
transform suitable for graphic processing units (GPU).  We use CUDA
and base our implementation on a {\sc Fortran90} routine included in a
publicly available parallel package, \s2hat.  We focus our attention
on the two major sequential steps involved in the transforms
computation, retaining the efficient parallel framework of the
original code.  We detail optimization techniques used to enhance the
performance of the CUDA-based code and contrast them with those
implemented in the {\sc Fortran90} version. We also present
performance comparisons of a single CPU plus GPU unit with the \s2hat
code running on either a single or 4 processors. In particular we find
that use of the latest generation of GPUs, such as NVIDIA GF100
(Fermi), can accelerate the spherical harmonic transforms by as much
as $18$ times with respect to \s2hat executed on one core, and by as
much as $5.5$ with respect to \s2hat on 4 cores, with the overall
performance being limited by the Fast Fourier transforms.

The work presented here has been performed in the context of the
Cosmic Microwave Background simulations and analysis. However, we
expect that the developed software will be of more general interest
and applicability. 
\end{abstract}

\section{Introduction}

Spherical harmonic transforms are ubiquitous  in diverse areas of science and practical applications, which need to deal with data distributed on a sphere.
In particular, they are heavily used in various areas of cosmology, such as studies of the cosmic microwave background (CMB) radiation and its anisotropies,
which have been our main motivations for this work.
CMB is an electromagnetic radiation left over after
the hot and very dense stage of early evolution of our Universe. The CMB measurements allow us to look back directly at  the Universe when its age
was only a small fraction ($\sim 3$\%) of its current one ($\sim13$Gyrs), and indirectly to learn about its status as far back as to $\sim10^{-35}$sec after its 
nominal beginning (so called Big Bang). 
Not surprisingly, the CMB measurements play a vital role in the present-day cosmology and have been a driving force behind turning it into a high precision, data-driven 
science it is today.

The CMB radiation is nearly isotropic but minute deviations, on order of $1$ part in $10^5$,  were first theoretically predicted and later detected. These  so-called anisotropies 
encode the information about the Universe, its past and composition, and their detection and characterization has the major target of the CMB observations
since the moment of its discovery in 1965.
Over the time progressively
more sophisticated
and advanced observational apparata have been designed and deployed in search for their  more subtle and taletelling characteristics.  These include three major CMB satellites 
 -- American: Cosmic Microwave Background Explorer (COBE)\cite{Smoot_etal_1992}, Wilkinson Microwave Anisotropy Probe (WMAP) \cite{Bennett_etal_2003}, and 
European Planck\footnote{http://sci.esa.int/science-e/www/area/index.cfm?fareaid=17} --  and a few  dozen of ground-based and balloon-borne projects. Some of these are operating at this time, including WMAP and Planck, and more are planned for the near and medium-term future, including potential new satellite missions, considered currently in Europe, US, and Japan.

The CMB detector technology has been improving quickly over the past decade, propelling a continuous increase of a number of detectors per experiment at the rate reminiscent
of the Moore's law. This in turn has been driving a similar increase of the CMB data sets sizes, posing a formidable challenge for the CMB data analysis.
This challenge can be only met if efficient numerical algorithms and the latest computer hardware are employed to match the data size increase
by a concurrent increase in our processing capability. Spherical harmonic transforms are some of  the most fundamental tools used in the CMB data processing.
 This is because the CMB signal is naturally a function of the observational direction and thus can be adequately described as  a field defined on a sphere. The spherical harmonic
 functions are a suitable basis to represent and manipulate such fields. The spherical harmonic transforms involve a decomposition of the signals defined on the sphere into a set of harmonic coefficients (i.e., a {\em direct} spherical harmonic transform) as well as synthesis 
 of the sky signal given a set of harmonic expansion coefficients (i.e, an {\em inverse} spherical harmonic transform). The latter is for instance
 a key step in massive Monte Carlo simulations used in the CMB data processing. As they usually require a very high resolution and precision, synthesis
 operations, referred hereafter as \alm2map transforms, are particularly time and resources consuming. In this work we therefore focus specifically on \alm2map
 transforms, and  discuss their implementation on the NVIDIA GPU architecture within the CUDA framework. Similar techniques to these described here should be sufficient
for an efficient implementation in this context of the direct {\tt map2alm} transforms. We leave those for the future work. We note that the spherical harmonic transforms
are commonly used beyond cosmology, for example, in geophysics, oceanography, or planetology and for all of which the implementation described here should be directly 
relevant.

There are several packages available implementing the spherical
harmonic transforms with {\sc healpix}\footnote{{\sc healpix}:
  http://healpix.jpl.nasa.gov/}, \s2hat\footnote{\s2hat:
  http://www.apc.univ-paris7.fr/$\sim$radek/s2hat.html}, {\sc
  GLESP}\footnote{GLESP: http://www.glesp.nbi.dk/}, {\sc
  ccSHT}\footnote{ccSHT: http://crd.lbl.gov/$\sim$cmc/ccSHTlib/doc/},
particularly popular in the CMB research.  Out of these, we have
selected \s2hat (Scalable Spherical Harmonic Transform) as the
starting point for this research and a reference for performance
comparisons. While all these packages implement similar numerical
algorithms, in particular \s2hat originated from the {\sc healpix}
routines, only \s2hat is not strictly tied to any specific sky
pixelization or discretization schemes.  \s2hat is written in {\sc
  Fortran90}, fully parallelized using MPI, and shows memory
scalability, good speedup and load-balance over a wide range of
considered problems.

\begin{figure}[htb]
  \centering
  {
		\includegraphics[width=0.45\textwidth]{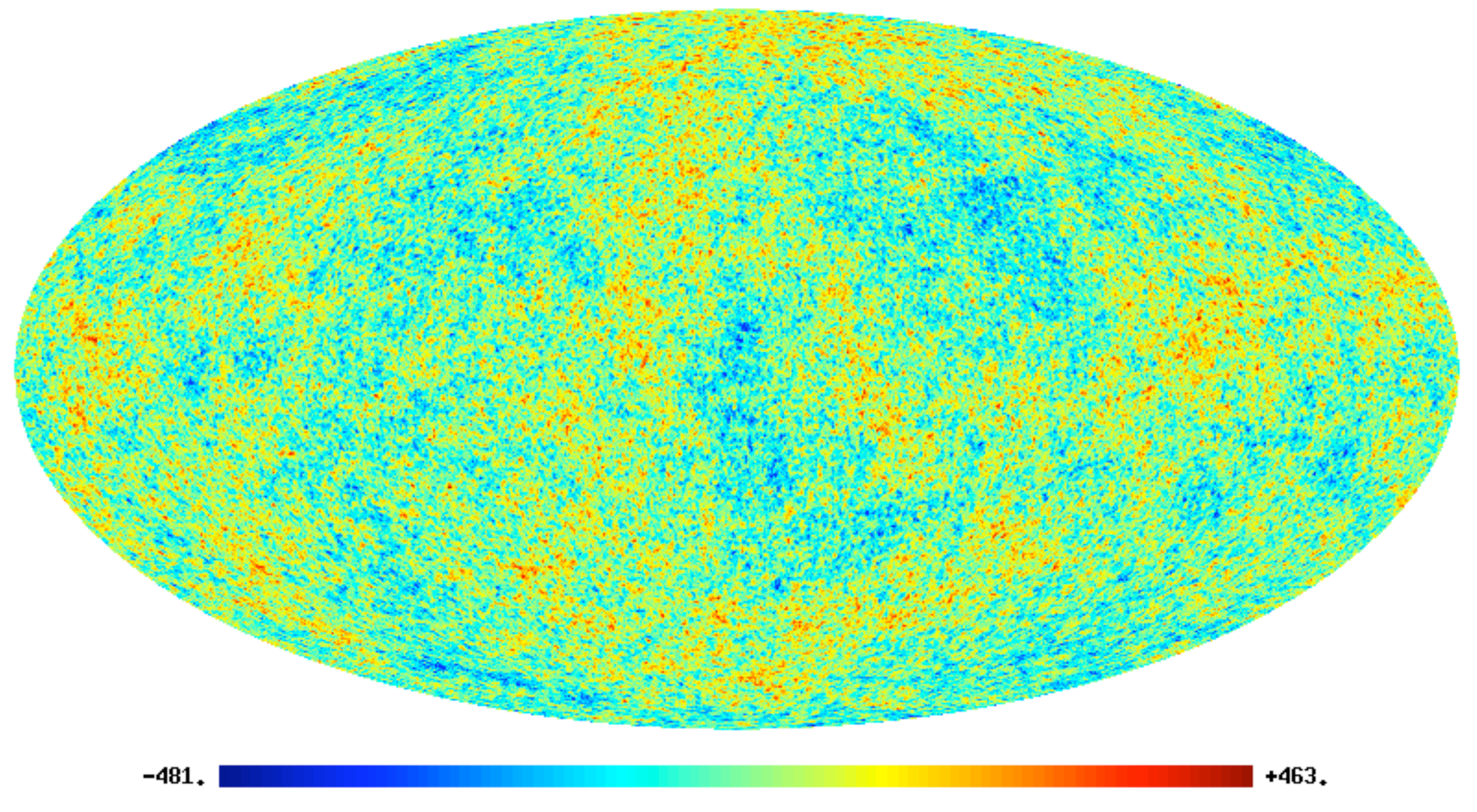}
  }
   \caption{An example output of the CUDA {\tt alm2map} routine the implementation of
   which is described in this paper. The figure shows a simulated picture of the sky
   as seen in the microwave band of electromagnetic radiation. The units are arbitrary.
  \label{fig:cmbSky}		
  }
\end{figure}

Our primary final target are however heterogeneous, multi-processor
systems made of multiple CPUs, each accompanied by a respective GPU.
As the first step towards achieving this goal we focus on porting the
two main, serial steps in the calculation of the transforms onto GPUs
and retain the data distribution layout and communication structure of
the original MPI code.  Consequently, when run on a
multi-processor/multi-GPU platform our code employs MPI calls to
distribute the data and workload over all the CPUs, which then send
them to their respective GPUs, where the bulk of the computation is
performed.  The MPI communication pattern and work distribution
inherited from the \s2hat package have been both demonstrated to scale
well in terms of memory load and execution time, and therefore we
anticipate that the overall performance gain due to the speed of the
GPUs will translate directly to an analogous total speed-up of the
complete code.  The proper end-to-end performance evaluation of the
code on heterogeneous systems is the object of our on-going work.  The
performance tests presented in this paper focus specifically on the
benefits due to GPUs and thus consider only single CPU/GPU cases.  In
addition this paper provides a first detailed description of the MPI
parallelization of the efficient, scalable spherical harmonic
transforms as implemented in \s2hat. Fig. ~\ref{fig:cmbSky} provides
an example result produced with our CUDA code. (An interested reader
can compare this figure with the actual observations produced by the
WMAP satellite and published in \cite{Bennett_etal_2003}).

The paper is organized as follows.  In section \ref{sec:spHarmonicTr} we introduce the spherical harmonic transforms and outline their MPI-based implementation.  After a brief introduction to the CUDA programming model in section \ref{sec:cuda}, we present a detailed description of our modified algorithm suitable for GPUs in section \ref{sec:alm2map_cuda} and associated optimizations in section \ref{sec:optimizations}.  Section \ref{sec:perf} present some comparison results of both implementations.  Section \ref{sec:concl} concludes the paper.

\section{Spherical harmonic transforms}
\label{sec:spHarmonicTr}

\subsection{Algebraic background}
For a real, scalar field, $\bm{s}$, defined on the ${\cal S}^2$-sphere the pair of  spherical harmonic transforms is defined as follows,
{\small
\begin{eqnarray}
\bm{a}_{\ell m} & = & \sum_{\l\{\theta_p,\phi_p\r\}}\,\bm{s}\l(\theta_p, \phi_p\r)\,Y_{\ell m}\l(\theta_p,\phi_p\r),
\label{eqn:map2almDef}
\\
\bm{s}\l(\theta_p,\phi_p\r) & = & \sum_{\ell=0}^{\ell_{max}}\,\sum_{m=-\ell}^{\ell}\,\bm{a}_{\ell m}\,Y_{\ell m}\l(\theta_p,\phi_p\r).
\label{eqn:alm2mapDef}
\end{eqnarray}
}
Here the coefficients $\bm{a}_{\ell m}$ define a harmonic representation of the field $\bm{s}$, $Y_{\ell m}$ stands for a spherical harmonic, and $\l(\theta, \phi\r)$
denote standard spherical coordinates. As in all practical application the field is pixelized or sampled on a discrete set of points, we have replaced the
continuous integral in Eq.~\ref{eqn:map2almDef} by a discretized summation, which now goes over all the pixels on the sky. The upper limit, $\ell_{max}$ in
Eq.~\ref{eqn:alm2mapDef} defines the band-limit of the field $\bm{s}$ and is considered to be finite. In the CMB application it is usually determined by an experiment
resolution and its typical values are $\ell_{max} = {\cal O}\l(10^3-10^4\r)$. Hereafter,
we will focus on the second of these two transforms and refer to it as the \alm2map transform. Its objective is to reconstruct, or synthesize, the field, $\bm{s}$, from its 
harmonic coefficients  $\bm{a}_{\ell m}$ on a grid of points $p$. (Hereafter, we will drop the index $p$ for shortness.)

The spherical harmonics are defined as,
{\small
\begin{eqnarray}
Y_{\ell m}\l(\theta, \phi\r) \equiv {\cal P}_{\ell m}\l(\cos \theta\r) e^{i m \phi}
\end{eqnarray}
}
where {\em renormalized} associated Legendre functions, ${\cal P}_{\ell m}\l(\cos \theta\r)$ are solutions of the Hemholtz equations, e.g., \cite{ArfkenBook},
 and their normalization is selected to ensure that $Y_{\ell m}$ constitute an orthonormal basis on the sphere.
{\small
\begin{eqnarray}
&& \bm{s}\l(\theta,\phi\r) =  \sum_{\ell=0}^{\ell_{max}}\,\sum_{m=-\ell}^{\ell}\,\bm{a}_{\ell m}\,Y_{\ell m}\l(\theta,\phi\r) \,  \nonumber \\
&=&  \, \sum_{\ell=0}^{\ell_{max}}\,\bm{a}_{\ell 0} \, {\cal P}_{\ell 0}\l(\cos\theta\r)
 +  \sum_{m=1}^{\ell_{max}}\, e^{im\phi} \, \sum_{\ell=m}^{\ell_{max}}\,\bm{a}_{\ell m}\,{\cal P}_{\ell m}\l(\cos\theta\r)  
 +
\sum_{m=1}^{\ell_{max}} \, e^{-i m\phi} \, \sum_{\ell=m}^{\ell_{max}}\,a_{\ell m}^\dagger\,{\cal P}_{\ell m}\l(\cos\theta\r),
\label{eqn:harmDecomp} 
\end{eqnarray}
}
where we use the fact that,
{\small
\begin{eqnarray}
{\cal P}_{\ell m}\l(\cos\theta\r) & = & \l(-1\r)^m\,{\cal P}_{\ell \l(-m\r)}\l(\cos\theta\r)\\
\bm{a}_{\ell m} & = & \l(-1\r)^m \, \bm{a}^\dagger_{\ell\l(-m\r)}.
\end{eqnarray}
}
The latter explicitly assumes that the map, $\bm{s}$, is real and a dagger denotes a complex conjugation.
We can introduce now  a set of functions, $\bm{\Delta}_m\l(\theta\r)$, such as,
{\small
\begin{equation}
\bm{\Delta}_m\l(\theta\r) \equiv 
\l\{ 
\begin{array}{l l}
{\displaystyle \sum_{\ell=0}^{\ell_{max}}\,\bm{a}_{\ell 0} \, {\cal P}_{\ell 0}\l(\cos\theta\r),} & {\displaystyle m = 0;}\\
{\displaystyle \sum_{\ell=m}^{\ell_{max}}\,\bm{a}_{\ell m}\,{\cal P}_{\ell m}\l(\cos\theta\r),} & {\displaystyle m > 0;}\\
{\displaystyle \sum_{\ell=\l|m\r|}^{\ell_{max}}\,\bm{a}_{\ell \l|m\r|}^\dagger\,{\cal P}_{\ell \l|m\r|}\l(\cos\theta\r),} & {\displaystyle m < 0,}
\end{array}
\r.
\label{eqn:DeltaDef}
\end{equation}
}
and rewrite Eq.~(\ref{eqn:harmDecomp}) as,
{\small
\begin{equation}
\bm{s}\l(\theta,\phi\r)  \, =  \, \sum_{m=-\ell_{max}}^{\ell_{max}} \, e^{i m \phi}\,\bm{\Delta}_m\l(\theta\r).
\label{eqn:Delta2map}
\end{equation}
}
The associated Legendre functions can be computed via a 2-point recurrence, e.g., \cite{ArfkenBook}, with respect to the multipole number, $\ell$. It reads,
{\small
\begin{eqnarray}
{\cal P}_{\ell+2, m}\l(x\r) = \beta_{\ell+2, m}\l[ x \,  {\cal P}_{\ell+1, m}\l(x\r) + \frac{1}{\beta_{\ell+1, m}} {\cal P}_{\ell m}\l(x\r)\r]
\label{eqn:assLegRec}
\end{eqnarray}
}
where
{\small
\begin{eqnarray}
\beta_{\ell m} = \sqrt{ \frac{4 \,\ell^2-1}{\ell^2-m^2}}.
\label{eqn:betaDef}
\end{eqnarray}
}
The recurrence is initialized by the starting values,
{\small
\begin{eqnarray}
{\cal P}_{m m}\l(x\r) & = & \frac{1}{2^m\,m!}\, \sqrt{\frac{\l(2m+1\r)!}{4\pi}}\,\l(1-x^2\r)^{m}
  \nonumber \\
  &\equiv& \mu_m \, \l(1-x^2\r)^{m},
\label{eqn:pmm}
\\
{\cal P}_{m+1, m}\l(x\r) & = & \beta_{\ell+1,m}\,x\,{\cal P}_{mm}\l(x\r).
\label{eqn:pm1m}
\end{eqnarray}
}
The recurrence is numerically stable but requires double precision and a care has to be taken to ensure it does not under- or overflows. We describe a relevant algorithm in the next Section.  Eqs.~\ref{eqn:DeltaDef} and~\ref{eqn:Delta2map} provide a basis for the numerical implementation of the spherical harmonic transforms.

\subsection{Current Approach}

A detailed description of the efficient serial implementation of the transforms can be found elsewhere \cite{Gorski_etal_2005}. Here we briefly outline the most important features, highlighting
in particular the parallel aspects.

\subsubsection{Numerical complexity}

From the sphere sampling considerations~\cite{Driscoll_Healy_1994}, we know that to properly sample a band-limited function with the band-limit set to $\ell_{max}$
we need roughly $n_{pix} \sim \ell_{max}^2$ points on the sphere. Therefore to perform the operations required to calculate $\Delta\l(\theta\r)$, and
as detailed in Eqs.~\ref{eqn:DeltaDef}, we need as many as ${\cal O}\l(n_{pix}^2\r)$ floating point operations (FLOPs). This is because for each of $n_{pix}$ pixels we have 
to do the ${\cal P}_{\ell m}$ recurrence for
all $\ell$ and $m$ numbers, and there are ${\cal O}\l(\ell_{max}^2\r) \sim {\cal O}\l(n_{pix}\r) $ of $\l(\ell, m\r)$ pairs  for a properly sample field.
This is clearly a prohibitive scaling.
It can however become more favorable if the problem is restricted to some specific sky pixelization/discretization schemes~\cite{Driscoll_Healy_1994}. In particular, in the following we will
always assume that all pixels/sky samples are arranged in a number of  so-called iso-latitudinal rings, each of which have the same value of the polar angle, $\theta$. Typically there will be
$n_{rings}\sim \ell_{max}$ rings with each ring uniformly sampled $n_{\phi}\sim \ell_{max}$ times. Moreover, we will assume that the sky is pixelized symmetrically with respect 
to the equator.
Such schemes indeed have been proposed and demonstrated to work well in practice \cite{Driscoll_Healy_1994, Muciaccia_etal_1997, Gorski_etal_2005, Doroshkevich_etal_2005} in a number of applications. 
With these constraints imposed on the pixelization the scaling for Eq.~\ref{eqn:DeltaDef} is now ${\cal O}(n_{pix}^{3/2})$, given that the full ${\cal P}_{\ell m}$ recurrence needs 
to be now done only ones 
for each of the rings. The numerical cost of the final summation, Eq.~\ref{eqn:Delta2map}, is then sub-dominant as it can be implemented using Fast Fourier transform (FFT) techniques, at the total cost of ${\cal O}\l(n_{pix}\ln n_{\phi}\r)$ FLOPs.
 
We note here
in passing that for this class of pixelizations even faster algorithms have been proposed with the complexity either on order of ${\cal O}[n_{pix}(\ln n_{pix})^2]$~\cite{Driscoll_Healy_1994} 
or ${\cal O}\l(n_{pix} \ln n_{pix}\r)$~\cite{Tygert2008}. However, they have a significant prefactor,  involve complex algorithmic solutions, and have not been demonstrated to be numerically 
viable for $\ell_{max} \gg 100$.

\subsubsection{Algorithm}

The implementation of Eqs.~\ref{eqn:DeltaDef} and~\ref{eqn:Delta2map} is rather straightforward. The pseudo code is outlined as Algorithm~\ref{algo:alm2mapBasic}.
\begin{algorithm}
\caption{{\sc Basic \alm2map algorithm}}
\label{algo:alm2mapBasic}
\begin{algorithmic}
{\small \sl
\STATE{
{\sc step 1 - $\bm{\Delta}_m$ calculation}
 \STATE{
  {\sc comment:} Algorithm~\ref{algo:assLegRec} has to be embedded below.
  }
}
  \FOR{every ring $r$} 
    \FOR{every $m = 0, ..., {m}_{max}$} 
      \FOR{every  $\ell = m, ..., \ell_{max}$} 
  \STATE{
  -- compute ${\cal P}_{\ell m}$ via the 2-point recurrence, Eq.~\ref{eqn:assLegRec};
  }
  \STATE{
  -- update $\bm{\Delta}_m\l( r\r)$, given input $\bm{a}_{\ell m}$ and computed ${\cal P}_{\ell m}$, Eq.~\ref{eqn:DeltaDef};
  }

  \ENDFOR{ ($\ell$)}
  
    \ENDFOR{ ($m$)}

\STATE{
{\sc step 2 - $\bm{s}$ calculation}
}
  
\STATE{
-- calculate $\bm{s}$ via FFT, given  $\bm{\Delta}_m\l(r\r)$ pre-calculated for all $m$;
}
 \ENDFOR{ ($r$)}
}
\end{algorithmic}
\end{algorithm}
Two steps which require somewhat more attention are the recurrence and the FFT. The two point recurrence as the one in Eq.~\ref{eqn:assLegRec} spans a huge dynamic range
of values. This range depends on the values of $\ell$, which need to be considered, but already for values  as low as ${\cal O}\l(10^2\r)$ it exceeds that 
accorded to a double precision number on a typical processor.  As we have freedom to rescale all the values of ${\cal P}_{\ell m}$, we can try to make use of it to rescale the recurrence starting values, e.g., ${\cal P}_{mm}$ and ${\cal P}_{m+1, m}$, appropriately to avoid the overflows later. However, this simple fix works only as long as the rescaled values of ${\cal P}_{mm}$ and ${\cal P}_{m+1, m}$ do not cause underflows. Though this on its own is not an issue, as these two will be typically set to zero what is clearly a good approximation to their values,  our 2-point recurrence as a result 
will never produce any other outcome than zero. This will apply also to  those Legendre functions, which initially produced the overflow and were supposed to be brought to within the representable range of values via the rescaling.
A more robust solution to the problem is that of real-time rescaling.  In the scheme usually used for this the newly computed values are tested if they 
approach over- or underflow limits and whenever this is the case they are rescaled appropriately. The rescaling coefficients (e.g., in form of their  logarithms) are kept tracked of
and used to rescale the computed values of ${\cal P}_{\ell m}$ at the end as required. This scheme is based on two facts. First,  that the values of the Legendre functions calculated
via the recurrence change gradually and rather slowly on each step. Second, that their final values are well-within the range of the double precision values.

The specific implementation of these ideas used in the \s2hat software, and derived from the solution coded in the {\sc healpix} package, uses a precomputed vector of
values,  sampling the dynamic range of the representable double precision numbers and thus avoids any explicit computation of numerically-expensive logarithms and exponentials. The scaling vector, referred to hereafter as a rescale table is used to compare the values of ${\cal P}_{lm}$ computed on each
step of the recurrence, and then used to rescale them if needed.  

The respective pseudo-code for the Legendre function recurrence is presented as Algorithm~\ref{algo:assLegRec}.
\begin{algorithm}
\caption{{\sc 2-point associated Legendre recurrence}}
\label{algo:assLegRec}
\begin{algorithmic}
{\small \sl
\STATE{
-- initialize the rescaling table;
}
\STATE{
-- precompute $\bm{\mu}$ coefficients, Eq.~\ref{eqn:pmm};
}
  \FOR{every ring $r$} 
    \FOR{every $m = 0, ..., {m}_{max}$} 
    \STATE{
    -- initialize the recurrence: ${\cal P}_{mm}$, ${\cal P}_{m+1, m}$, Eqs.~\ref{eqn:pmm}~\&~\ref{eqn:pm1m}, using precomputed $\bm{\mu}_m$;
     }
     \STATE{
     -- precompute recurrence coefficients, $\beta_{\ell m}$ (fixed $m$, $\ell \in\l[m,\ell_{max}\r]$), Eq.~\ref{eqn:betaDef};
     }
      \FOR{every  $\ell = m+2, ..., \ell_{max}$} 
  \STATE{
  -- compute ${\cal P}_{\ell, m}$ given ${\cal P}_{\ell-1, m}$ and ${\cal P}_{\ell-2 m}$, given precomputed $\bm{\beta}_{\ell m}$, Eq.~\ref{eqn:assLegRec};
  }
  \STATE{
  -- test the value of  ${\cal P}_{\ell+2 m}$ against the rescaling table;
  }
   \STATE{
  -- rescale ${\cal P}_{\ell+2, m}$ and ${\cal P}_{\ell+1, m}$ if needed, keep the info about the rescaling coefficient;
  }
  \STATE{
  {\sc comment:} ${\cal P}_{\ell m}$ needs to be scaled back before being used in the calculations of the functions, $\bm{\Delta}_m$;
  }

  \ENDFOR{ ($\ell$)}
    \ENDFOR{ ($m$)}
      \ENDFOR{ ($r$)}
  
}
\end{algorithmic}
\end{algorithm}
The associated Legendre function recurrence is normally performed on-the-fly and Algorithm~\ref{algo:assLegRec} is thus merged with the algorithm for the \alm2map
transform, Algorithm~\ref{algo:alm2mapBasic}.

The application of the FFTs in the last step of Algorithm~\ref{algo:alm2mapBasic} also requires some care. This is because the number of samples per ring
may be either  larger or smaller than the number of available $m$-modes. The former case can be dealt with by assuming the missing mode amplitudes to be zero. In the
latter case the extra $m$ modes have to be wrapped up and co-added to the modes present in the box. We refer the reader to paper \cite{Gorski_etal_2005}  
for more details of the involved calculations.

\subsubsection{MPI parallelism}

The structure of the parallel implementation of Algorithm~\ref{algo:alm2mapBasic} is determined by the data layout. For a properly sampled full sphere the input set of 
the harmonic coefficients, $\bm{a}_{\ell m}$, and the output sky map, $\bm{s}$,  are roughly of the same size as $\sim n_{pix}\sim \ell_{max}$. These two objects are typically
large and preferably have to be distributed.

 \s2hat divides the 2-dimensional
$\bm{a}_{\ell m}$ array by assigning to a processor $i$ a subset of all coefficients with predefined $m$ values, ${\cal M}_i$. The 2-dimensional sky map $\bm{s}$ is treated as a collection of rings, $r$, so
a subset, ${\cal R}_i$ of complete rings is distributed to a processor $i$.  Recall that $r$ corresponds to a subset of pixels, identified by a unique $\theta$.  This ensures the memory scalability of the algorithm if only the sizes of all the sets, ${\cal M}_i$ and
${\cal R}_i$, decrease as roughly $\sim 1/n_{procs}$. 

The algorithm, see Algorithm~\ref{algo:alm2mapS2HAT}, proceeds then in two steps. First, given the input subset of all $\l\{\bm{a}_{\ell m}, m \in {\cal M}_i\r\}$ coefficients, each processor calculates, using Eq.~\ref{eqn:DeltaDef}, the $\bm{\Delta}_m$ functions for {\em every} ring, $r$, of the sphere and $m \in {\cal M}_i$. Once this is done, the global communication is performed
so that at the end each processor has in its memory $\bm{\Delta}_m$ functions calculated for the subset of rings as assigned to this processor, ${\cal R}_i$, and {\em all} $m$ values.
Given these data each processor performs the second step of the sky calculations, Eq.~\ref{eqn:Delta2map}, using FFTs. We note that once the data distribution is performed
then steps 1 and 2 of the algorithm are embarrassingly parallel. The memory required to store the intermediate products, $\bm{\Delta}_m$, is on order of ${\cal O}\l(n_{pix}/n_{procs}\r)$ 
and therefore are comparable to that used to store the input and output objects in their distributed form. Also like the latter they decrease as a number of employed processors increases, preserving the overall memory-scalability of the algorithm.
\begin{algorithm}
\caption{{\sc \s2hat \alm2map algorithm - mpi implementation}}
\label{algo:alm2mapS2HAT}
\begin{algorithmic}
\medskip
{\small \sl
  \STATE{
  {\sc comment:} Code executed by each MPI process.
  }

\STATE{\sc step 1 - $\bm{\Delta}_m$ calculation}\\
\STATE{
-- {\sc step 1.1} - initialize the rescaling table;
}
\STATE{{\sc step 1.2} - precompute $\bm{\mu}$ coefficients, Eq.~\ref{eqn:pmm};}\\
  \FOR{\underline{every} ring $r$} 
    \FOR{every $m \in {\cal M}_i$:} 
     \STATE{
    -- {\sc step 1.3} - initialize the recurrence: ${\cal P}_{mm}$, ${\cal P}_{m+1, m}$, Eqs.~\ref{eqn:pmm}~\&~\ref{eqn:pm1m}, using precomputed $\bm{\mu}_m$;
     }
    \STATE{
      -- {\sc step 1.4} - precompute recurrence coefficients, $\beta_{\ell m}$ (fixed $m$, $\ell \in\l[m,\ell_{max}\r]$), Eq.~\ref{eqn:betaDef};
     }
      \FOR{every  $\ell = m+2, ..., \ell_{max}$} 
  \STATE{
  -- {\sc step 1.5} -- compute ${\cal P}_{\ell m}$ via the 2-point recurrence, given precomputed $\bm{\beta}_{\ell m}$, Eq.~\ref{eqn:assLegRec};
   }
    \STATE{
  -- {\sc step 1.6} - test the value of  ${\cal P}_{\ell m}$ against the rescaling table;
  }
    \STATE{
    -- {\sc step 1.7} - update $\bm{\Delta}_m\l( r\r)$, given input $\bm{a}_{\ell m}$ and computed ${\cal P}_{\ell m}$, Eq.~\ref{eqn:DeltaDef};
  }

  \ENDFOR{ ($\ell$)}

    \ENDFOR{ ($m$)}
      \ENDFOR{ ($r$)}
\medskip
\STATE{{\sc global communication}}\\
\STATE{
-- redistribute $\l\{\bm{\Delta}_{m}\l(r\r),\; m\in {\cal M}_i,\; \hbox{{\rm all $r$}}\r\} {\hbox{ {\tt \scriptsize MPI\_Alltoallv}}\atop\longrightarrow} \l\{\bm{\Delta}_{m}\l(r\r),\; r\in {\cal R}_i,\; \hbox{{\rm all $m$}}\r\}$
}
\medskip
\STATE{\sc step 2 - $\bm{s}$ calculation}\\
\FOR{every ring $r \in {\cal R}_i$}
\STATE{
-- using FFT calculate $\bm{s}\l(r, \phi\r)$ for  all samples $\phi \in r$,  given pre-computed $\bm{\Delta}_m\l(r\r)$ for  \underline{all} $m$.
}
\ENDFOR{ ($r$)}
}
\end{algorithmic}
\end{algorithm}
In its current form the \s2hat implementation is memory-distributed and implemented using MPI. Adding the openMP layer could be certainly useful as it could help
to alleviate the communication bottleneck and thus facilitate running the library at even higher concurrencies. Nevertheless openMP is expected to have no
major impact on the computational efficiency of steps 1 and 2 of the algorithm due to their embarrassingly parallel character. Considering the accelerators, such as 
GPUs, is therefore a potentially attractive avenue to explore.

\section{NVIDIA CUDA Programming Model}
\label{sec:cuda}

NVIDIA CUDA is a general purpose parallel computing architecture - with a new parallel programming model and instruction set architecture - that takes advantage of the parallel compute engine in NVIDIA GPUs.  CUDA comes with a software environment that allows developers to use C as a high-level programming language. Other languages and application programming interfaces are also supported.

A CUDA-enabled GPU is basically a manycore chip consisting of hundreds of simple cores, called Streaming Processors (SP), together with control logic for the different levels of encapsulation. Each SP executes instructions sequentially, has a pipeline, 2 arithmetical-logical units and one floating point unit. The older generations do not have a cache.  Several SPs are encapsulated in a Streaming Multiprocessor (SM). Multiple SMs form a Texture/Processor Cluster (TPC). All the TPCs form the Streaming Processor Array.

Inside one SM, the SPs execute in a SIMD fashion, while different SMs may execute different parts of the instruction stream in a SPMD fashion. Each SM also manages hundreds of active threads in a cyclic pipeline in order to hide memory latency. In practice, 32 threads are grouped together and scheduled as a Warp, executing the same instructions.

The latest two architecture generations available from NVIDIA are GT200 and GF100 (Fermi). The best model of GT200 has a total of 240 SPs (8 SP/SM x 3 SM/TPC x 10 TPC), with 16 KB of shared memory per SM and 16K available registers per block. It has a better implementation for memory fetching, reducing the earlier generation (G80) bottleneck produced by uncoalesced read/writes. Its theoretical peak performance is 1062.72 GFLOPS, in single precision. For double precision, FLOP count is 8 times lower. GF100 increases the number of SPs to 512 (32 SP/SM x 4 SM/GPC x 4 GPC), shared memory to 64 KB and adds a Level 1 caching mechanism. Theoretical peak performance for single precision is 1344.96 GFLOPS and double precision is half this value. However, for the commercial GTX 480 graphics card, the double precision performance is intentionally limited to one eighth of single precision. The only products using the entire DP capacity of the chip are those in the NVIDIA Tesla line.

The CUDA threads are grouped together into a series of thread blocks. All threads in a thread block execute on the same SM, and can synchronize and exchange data using shared memory. Synchronization between thread blocks is not possible. One more level of encapsulation is possible as the thread blocks can be grouped in independent grids. 
The Fermi chip is equipped with a resizable Level 1 cache (using shared memory for storage) which has the purpose of removing manual data copy between the slow device global memory and the fast shared memory. It is enabled by default, but its size can be switched between 16 KB to 48 KB. 

Various limitations and roadblocks have to be taken into account while developing algorithms on such architectures. First, the performance for  
the total chain of computation has to take into account the transfer time between the CPU and GPU main memory. For smaller algorithms, this transfer
can represent more than ten times the computation step itself. One is then tempted to fit large segments of data in the GPU memory to limit
the number of times the data has to be sent and received. However, doing so will put a heavy constraint over the memory management code inside 
the kernel. To fasten access to the global memory, data is usually fetched into some local memory segments. However, using large segments of local 
memory reduces the size of the registers bank, thus leading to a large amount of slower code spill. Balancing between transfer time, size and actual
memory management strategy is a very important problem to solve for any high performance algorithm executing on GPUs.  We address these issues in the following section in the context of our application. 

We also considered OpenCL as alternative for the development, but opted for CUDA, which we expect is best  suited to exploit the capabilities of the studied hardware.

\section{\alm2map with CUDA}
\label{sec:alm2map_cuda}

Programming philosophy for CUDA dictates using fine grained parallelism and launching a very large number of threads in order to use all the available cores and hide memory latency. Since the loop computing the two-point recurrence is serial in nature and therefore cannot be broken into parallel segments, there are two remaining choices for parallelization: the $m$-loop and the ring loop.

The initial CPU approach involved parallelizing only the $m$-loop, by having each process compute all the ring values for a subset of $m$ values. This method of parallelization makes it easy to write code for MPI, as each process works on a subset of $m$ values.

This approach is not appropriate for the GPU however, because of shared memory limitations. The size of vector \blm{}, Eq.~\ref{eqn:assLegRec}, depends on $\ell_{max}$ and therefore cannot be stored in shared memory. Its values need to be recomputed for each $m$ and are accessed sequentially in the $\ell$-loop. A possible implementation would require re-computing the \blm{} coefficients for each pass through the $\ell$-loop. However, these expensive, repeating calculations would seriously limit performance.

\begin{figure}[htb]
  \centering
  \subfloat[Access pattern for input data ( \aalm{} )]
  {
  	\centering
		\includegraphics[width=0.35\textwidth]{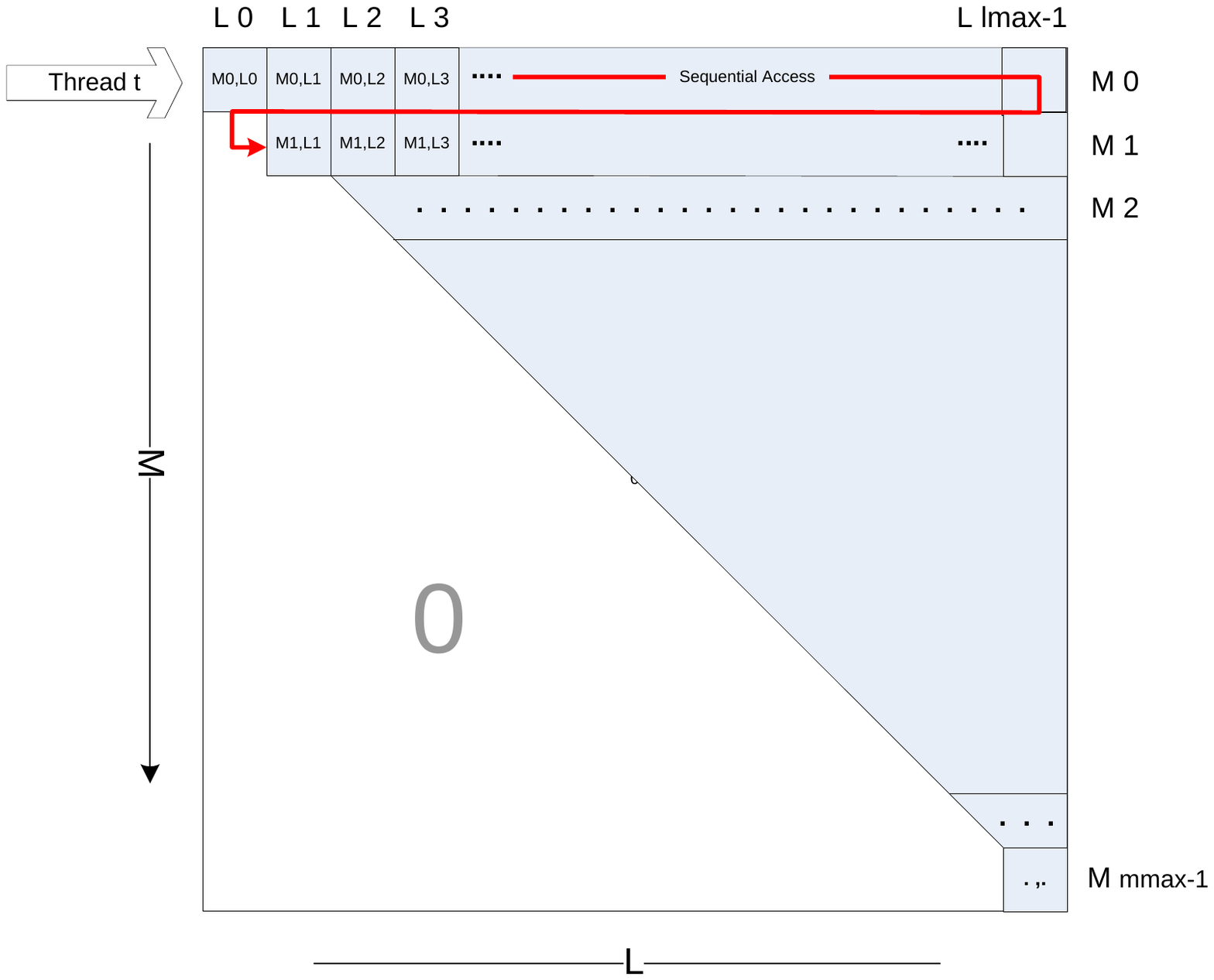}
		\label{fig:input}		
	}
  \hspace{0.2cm}
  \subfloat[Output matrix (\deltam) write pattern]
  {
		\includegraphics[width=0.3\textwidth]{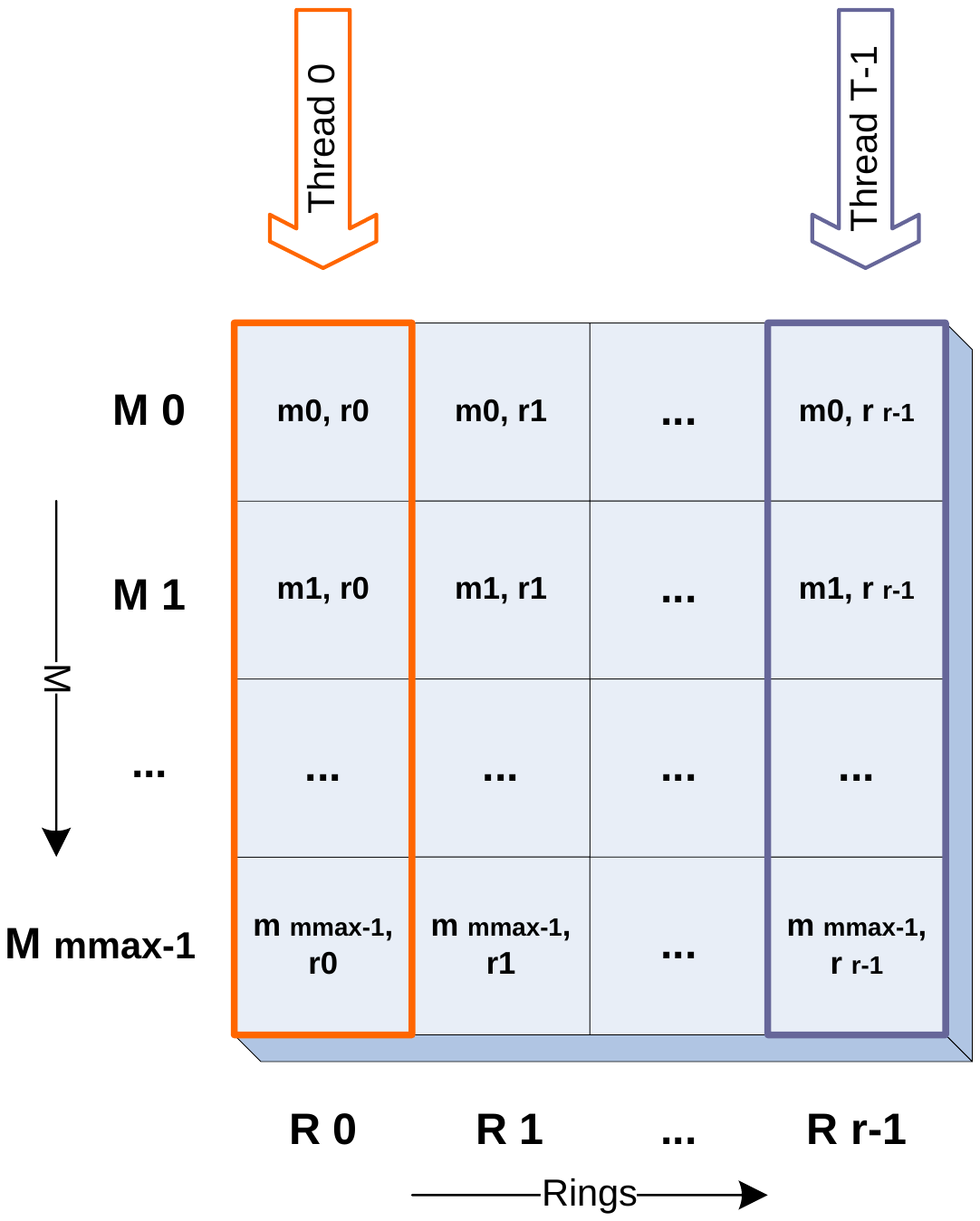}
		\label{fig:output}		
	}
	\caption{Input and output data access patterns}
\label{fig:inputoutput}
\end{figure}

Parallelizing the ring loop (step 1.1 in algorithm \ref{algo:alm2mapS2HAT_CUDA}) avoids this problem and has additional advantages. Each thread is assigned a number of rings for which it computes the 2-point recurrence for all $m$-values. The consequence is that each thread processes \aalm{} values at the same $m$ and $\ell$ coordinates, in parallel. This makes it easy to plan the computation of \blm{} and \miu{}, Eq.~\ref{eqn:pmm}, in segments, as well as caching the \aalm{} values. An important added benefit is reusing these two vectors, by sharing them inside a thread block.
\begin{algorithm}
\caption{{\sc \s2hat \alm2map  algorithm - cuda implementation}}
\label{algo:alm2mapS2HAT_CUDA}
\begin{algorithmic}
\medskip
{\small \sl

	\STATE{\sc step 1 - $\bm{\Delta}_m$ calculation}\\
	\medskip
	\STATE{	-- {\sc step 1.1} - assign rings for each \underline{thread}						}	
	
	\FOR{every $r \in {\cal R}_j$}
		\FOR{every $m \in {\cal M}_i$} 
	
			\STATE{  -- \textsc{step 1.2} - thread 0 in block computes a segment of \miu{};	}
			  
			\FOR{every  $\ell = m+2, ..., \ell_{max}$}
			
	       \STATE{  -- \textsc{step 1.3} - use precomputed or, if needed, precompute in parallel a segment of \blm{}, Eq.~\ref{eqn:betaDef};						}
	       \STATE{  -- \textsc{step 1.4} - use fetched or, if needed, fetch in parallel a segment of \aalm{} map data; 	}

		     \STATE{  -- \textsc{step 1.5} - compute ${\cal P}_{\ell m}$ via the 2-point recurrence, Eq.~\ref{eqn:assLegRec};	}

		     \STATE{  -- \textsc{step 1.6} - handle overflow/underflow using rescaling table;					}
	  	
	  	\STATE{  -- \textsc{step 1.7} - update $\bm{\Delta}_m\l( r\r)$, given prefetched $\bm{a}_{\ell m}$ and computed ${\cal P}_{\ell m}$, Eq.~\ref{eqn:DeltaDef};													
		}
	
	       \ENDFOR{ ($\ell$)}	
	
		\ENDFOR{ ($m$)}
	\ENDFOR{ ($r$)}
	      
\medskip
\STATE{\sc global communication}\\
\STATE{
-- redistribute $\l\{\bm{\Delta}_{m}\l(r\r),\; m\in {\cal M}_i,\; \hbox{{\rm all $r$}}\r\} {\hbox{ {\tt \scriptsize MPI\_Alltoallv}}\atop\longrightarrow} \l\{\bm{\Delta}_{m}\l(r\r),\; r\in {\cal R}_i,\; \hbox{{\rm all $m$}}\r\}$ 
}
\medskip
\STATE{\sc step 2}\\
\STATE{
-- using FFT calculate $\bm{s}\l(\theta, \phi\r)$ for  all samples for every, $r \in {\cal R}_i$,  given pre-computed $\bm{\Delta}_m\l(\theta\r)$ for $r\in{\cal R}_i$ and \underline{all\phantom{y}} $m$.}
}
\end{algorithmic}
\end{algorithm}
Algorithm \ref{algo:alm2mapS2HAT_CUDA} shows the outline of the GPU computing kernel. It is observable that the three new steps (1.2, 1.3 and 1.4) are designed to work around the high latency device memory and take advantage of the fast, but small, shared memory. Steps 1.2 and 1.3 calculate the values of the \miu{} and \blm{} vectors in segments, as they do not fit in shared memory and it would be slow and wasteful to store them in global memory. Step 1.4 tries to keep a supply of \aalm{} values for the 2-point recurrence, therefore allowing a more continuous operation of the floating point units by decreasing memory wait time. Step 1.1 is where the threads select the rings on which to work upon. Since the $m$-loop and ring-loops are interchangeable, unlike the CPU version, the ring loop is first, allowing the sharing of the \miu{} and \blm{} vectors.  Figure \ref{fig:inputoutput} displays the acces by each thread of the 2-dimensional arrays \aalm{} and \blm{}. 

\section{Optimizations for GPU}
\label{sec:optimizations}

GPU code optimization uses different rules than regular, CPU based code optimization. Due to the massively parallel structure of the target architecture,
GPU code needs to fulfill a different set of constraints. Among these, the relationship between the cost of memory access and
amount of computations per kernel is exacerbated. As shown in \cite{Volkov08} for example, it can be far more beneficial to recompute
large segments of constant values instead of fetching them from main memory. Others \cite{Nukada09} show that, in some cases, the most direct algorithm can outperform
the CPU optimized one. Another source of performance loss is thread divergence due to asymmetrical branching in control flow. Such
divergence is usually detected by various profilers, but can prove hard to remove.

Based on guidelines for CUDA kernel optimization \cite{CUDAManual, CUDABestPractices, CUDAFermi} and previous experiences, we mainly looked at limiting the effect of the slow global memory by buffering, precalculating or reusing data, removing branching in performance-critical sections and canceling warp serialization.

\textbf{Array segmentation.}
Due to shared memory small size, it is required to compute the \blm{} vector in segments, on the fly  (step 1.3). Pre-calculating it entirely in device memory (akin to the original CPU implementation) would be very slow, as completing one $P_{\ell m}$ value requires reading the entire vector.
\blm{} segments are computed inside the $\ell$-loop. Since \blm{} is accessed sequentially, a portion of the vector is computed then used in the following steps of the recurrence. When existing values are exhausted, the next portion is computed. The size of the segment influences code performance, as it can be seen in the performance section.
The same philosophy is applied for the \miu{} vector. Only difference is the segments are computed inside the $m$-loop  (step 1.2).
The advantage of having the code process the same \aalm{} data is that the two vectors are computed only once (in a parallel and serial fashion, respectively) and then reused by all threads in a thread block. As expected, the runtime decreases with increase in the number of threads. 

A similar approach to segmentation is employed for offsetting memory latency for reading the \aalm{} coefficients and transferring them only once before being used by all threads in a block (step 1.4). \aalm{} values are transferred in segments during the Plm computation in step 1.5. Optimal segment size for all three vectors is found through testing. It is input size and platform dependent. Because of this, an autotuner is the best solution for obtaining the best possible performance. By running it for an input of the size targeted for computation, with all representative segment size and thread count variations, the best set of segment sizes can be selected. Currently, combinations that provide acceptable performance in all cases have been found by manual inspection and are being used as defaults in the code.

The nature of \blm{} and \aalm{} allows their values to be obtained in parallel, by computing or fetching (steps 1.3 and 1.4, respectively). The number of threads which perform this operation is directly linked to segment size. In particular, the segment size must be a multiple of the number of threads. This avoids additional code for handling outlier indexes in  performance-critical sections. Keeping in line with the CUDA guidelines on shared memory access for avoiding bank conflicts, the threads in a block calculate values sequentially, with a stride of block size. Due to its serial nature, \miu{} is computed by a single thread, while others wait for its completion (step 1.2).

There is one more type of global information, called pixelization data. It also comes in the form of two arrays, but they are not cached. This is because they are rarely accessed and caching would complicate the code with no speed benefits.

\textbf{Branch collapsing.}
Code branching can severely impair the performance of GPU code, as divergent code is executed sequentially, effectively canceling parallelism. For example, when an "if" statement is encountered, some threads execute the true branch, while the other wait, then execute the false branch with the first group waiting.
This problem is solved by collapsing the branch into code that has the same outcome as before, but can be executed in parallel by all threads. The computational overhead is smaller than that incurred by process-and-wait execution. Conditional assignments like \verb|if (c) v=tv else v=fv |are converted to \verb#v=c | tv & !c | fv#.  The use of binary operators makes this expression very fast to compute. It is a common technique when converting code to SIMD operations.
On the GPU however, this can be applied only on operations with integer numbers, as binary operators are not applicable to floating point operands. An equivalent version is based on multiplications and subtraction: \verb|v = tv*c + fv*(1-c)|. This severely increases the overhead and makes it applicable only in some cases.
For \s2hat code, this version was employed in both full and short form (if-then) resulting in decreased branching but with limited influence on execution time.


Other approaches have been tried for using the resources of the GPU as much as possible. While none of them provide increased performance, they do offer some insight into the operation of this new platform and serve as lessons for the future.  We describe them briefly in the following. 

\textbf{Warp serialization.} 
Warp serialization for arrays of double precision floating point stored in shared memory is a problem for GT200 chips. Since a memory bank holds only 32 bit values, a 64 bit value is stored in two different banks. When the number of threads grows beyond half the number of banks, values are accessed concurrently from the same bank. However, a bank can only service one request at a time and threads making additional requests are serialized, waiting for the previous request to complete. NVIDIA Programming guide suggests one possible solution as splitting the 64 bit value into 2 32-bit ones, storing them in two different vectors and then rejoining at use \cite[p. 156]{CUDAManual}. However, after testing this technique on each vector of double precision data stored in shared memory, it proved useless. On the GT200 architecture, the computational cost of splitting and joining the values outweighed that of warp serialization. Moreover, the newer GT400 chips addressed this problem and 64 bit floats no longer cause warp serialization.

\textbf{Dedicated scaling table.}
The scaling table is subject to a different kind of warp serialization. When threads in a block enter the rescaling phase, they access the data inside the array in a random fashion (some elements accessed by a single thread, others by multiple). Being small in size (21 64-bit values), the simplest approach for canceling serialization is having a copy of each table for each thread. However, experiments showed that while serialization does not occur, the time gain is insignificant even for small inputs. Also, as the number of threads increases, the amount of shared memory used becomes a limiting factor (for just 64 threads, 10.5 KB are needed).

\textbf{\blm{} precalculation.}
Based on the ability to execute a very large number of mathematical operations and the drawback of high device memory latency, a method for obtaining a good throughput is computing values on-the-fly instead of precalculating them. This trades computing cycles for memory cycles and some algorithms gained significant performance in this manner. \blm{} calculation inside the $\ell$-loop turned out to greatly increase computation time over both precalculation-based version and segment-based version. This is due to the high number of expensive operations involved in computing a single value of \blm{} (multiplications, divisions and square roots), making reuse essential. Computing the scaling factors at usage-time had the same problem of expensive operations (powers, in particular). 

\textbf{Branch collapsing for scaling.}
Each iteration of the inner, $\ell$-loop involves checking the value to be inside a validity interval and apply a scaling factor if this is not the case. This requires two "if" checks and can result in branching, impairing performance, as threads can go on 3 different execution paths. By collapsing the code using the technique described above, branching was reduced, but had the adverse effect of increased execution time. This is caused by the high number of multiplications forced on each thread by both the scaling code (which does not always execute) and operations introduced by the collapse method itself, since bitwise operators are not available for floating point.

\section{Experimental results}
\label{sec:perf}

Two platforms have been used for testing the code: GTX 260 for NVIDIA GT200 architecture and GTX 480 for the new NVIDIA GF100 (Fermi). Their host systems are: AMD Phenom 9850 (4 cores) with 8 GB of PC3200 DDR2 memory running on a MSI MS-7376 motherboard and Intel Core i7-960 CPU (4 cores, 8 processes with Hyperthreading) with 8 GB of PC3200 DDR2 memory running on a Gigabyte EX58-UD5, respectively.

The number of theoretical double precision FLOPS is significantly in favor of the GPUs, with a ratio of 1:2.2 and 1:3.2, respectively, when compared to the 51.2 GFLOPS double precision performance of Intel i7-960. The GPU FLOPS counts a FMADD operation as two separate ones, for an easier comparison with the CPU.
It is also taken into account the fact that the Fermi chip can process a FMADD and ADD operation in parallel.

Also, the GPU architecture's massive parallelism of 260 and 480 SPs indicates a large advantage over the 4 physical cores of the referece CPUs. Even though the algorithm is known for near-linear scaling, due to the memory bound nature of the code (obtaining 10-15\% of the theoretical peak performance), the algorithm was expected to get a significant, but limited, runtime improvement. It was anticipated that the high latency of the GPU global memory would further stall execution.

On the CPU, the Fortran algorithm was used as reference. Its efficiency was computed using the FLOP count returned by the PAPI package.

For the GPU, the execution time is calculated using the \texttt{gettimeofday()} library call between kernel launch and result retrieval. Because the consumer-grade cards used have limited memory, the largest dataset used is 4096x4096 and 5120x5120, respectively. In order to evaluate the possible runtime improvement for larger datasets, the output arrays were no longer allocated, allowing the input to fill the entire card memory. Results were written in a very small buffer (one value per thread), in order to maintain memory access and not distort the results. In this manner, the dataset limit was pushed to 9216x9216.

\subsection{Parameter setup for \deltam}

Algorithm performance is dependent on the following parameters: number of threads in a block, number of rings per thread, size of the buffers used for computing the \miu{} and \blm{} values, size of the input buffer, usage of 64-bit floating point split into 2 32 bit integers for storage in shared memory, usage of L1 cache (for Fermi) and the usage of a shared or dedicated rescaling table for each block. We tested each of these parameters for their influence on runtime and selected the best results.

Since computing the values for each ring can be done independently, each thread can calculate the values for any given number of rings. This would allow running a fixed number of threads, for a theoretical performance improvement (for example, running a number of threads equal to that of the SPs, resulting in no overhead from thread context switching -- \cite{Volkov08}). However, testing has shown that the optimum amount of rings for a thread is 1. Higher numbers result in significant performance degradation for any combination of parameters. Since other values would always return sub-par (and therefore useless) runtimes, all subsequent tests are performed with just one ring per thread.  

One of the difficulties raised by the GT200 architecture is the method used for storing 64 bit values in the shared memory banks. When the number of threads grows beyond half the number of banks, values are accessed concurrently from the same bank, triggering serialization which results in latency for data read/write. The solution for avoiding it employs a workaround method, suggested in the NVIDIA Programming Guide, by splitting the data into two 32 bit values, storing them in shared memory then merging back to the original form at retrieval. This is no longer necessary on the Fermi, as 64 bit array access no longer generates warp serialization.
This technique was tested on the \miu{}, \blm{} and \aalm{} arrays, in all their possible combinations (with or without using the L1 cache) as well as with all thread counts. It resulted in overall performance degradation for all cases. The reason for this is that the overhead for splitting and merging the values is greater than the time gained by avoiding serialization. The following tests were performed with 32-bit splitting disabled in all cases.

\begin{figure}
	\centering
		\includegraphics[scale=0.7]{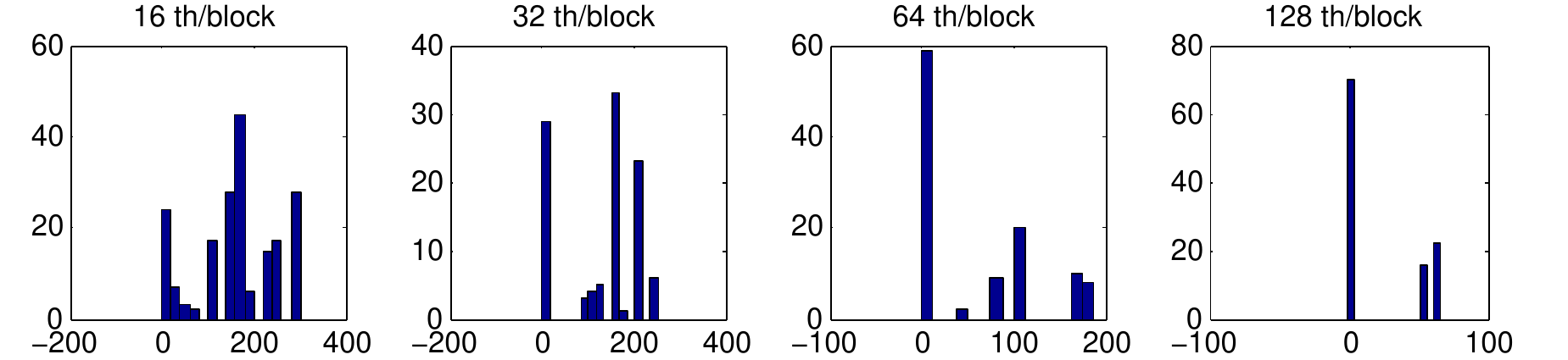}
	\caption{Histograms showing performance decrease (in percentages) due to L1 cache enabling on GTX 480}
	\label{fig:l1_cache}
\end{figure}

Another aspect tested was the cache system of the Fermi chip. This is designed to handle automatic fetching of values from device memory and store them into the fast shared memory, a task usually performed by hand by the CUDA programmer. Since array data is accessed sequentially (see figure \ref{fig:input}), caching should be straightforward and benefit from a hardware implementation. However, after running tests with all combinations of segment sizes, thread count per block and L1 cache on or off, it was determined that, for this particular algorithm, the caching implemented by the GPU never improves and often degrades performance. This is visible in figure \ref{fig:l1_cache}, which contains the histograms for performance degradation values when enabling L1 preference, for the different thread counts used. This decrease is most apparent for low thread counts (average speed degradation of 200\%). For 64 threads per block, activating L1-preference has no impact in half of the tests and ranges between 50\% and 200\% runtime increase for the rest. At 128 threads per block, half of the tests are not influenced, whilst the other suffer a 50-60\% speed decrease. The input size used is large, 4096x4096, in order to obtain a relevant result. We can conclude that, for this particular case, manual buffering outperforms the L1 cache system of the Fermi.

The rescaling table is the read only array most accessed by all threads. Each time an overflow or underflow event takes place inside the L-loop, a value in the table is read (algorithm \ref{algo:alm2mapS2HAT_CUDA} - step 1.6). Therefore, optimizing its access can have a significant impact on overall algorithm performance. Due to its small size (21 64 bit values), giving a dedicated copy of the table to each thread is a good way of avoiding warp serialization triggered by concurrent random access. However, as the number of threads increases, shared memory becomes a limiting factor. This is especially true on the GTX 280, which has just 16 KB available. Running the algorithm with the rescaling table shared by the thread block has shown that, contrary to expectation, the algorithm is faster by 30-50\%. 

Subsequent experiments used a shared rescaling table, had L1 cache preference (when ran on the Fermi) and 32-bit splitting disabled as well as treated just one ring per thread. The last three parameters to test are the lengths of the \miu{}, \blm{} and \aalm{} buffers. As expected, they have a major impact on the overall algorithm performance. Final performance values have been obtained by running all segment length combinations (16, 64, 128, 256 and 512 elements) with thread counts per block (16, 32, 64, 128, 256 and 512) for all input sizes. Since, for most cases, $m_{max}$ equals $\ell_{max}$ (the \aalm{} matrix is square), the data sets used as input use the same restriction.

Having such a large set of experimental data, analysis on the influence of each parameter was attempted, since it was not freely observable. It was revealed that segment length and thread count have a direct influence on the runtime, but a correlation between their association and runtime was not found. In addition, combinations that give the best results for a certain input size do not keep that property for other datasets. Also, it was discovered that the runtimes can be split into clusters of values, as defined by different parameter combinations. However, this grouping differs with the input size.

In order to obtain the best performance for any input size, an autotuner is the best solution. Combinations providing near-best runtimes have been found by manually taking the parameters that provide the best time for a certain dataset and analyzing its performance when applied for the other datasets. Out of these parameter sets, the best overall was selected and used as default. As sufficient experimental data is available, this process can be further refined.

By executing the algorithm on the massively parallel chip that is the GPU, the number of running threads and their configuration relative to the processing units influences runtime considerably. In order to produce the final performance results, the best time is selected when varying both the thread count in a block as well as the segment length.

\begin{figure}[htb]
  \centering
  \subfloat[GTX 260]
  {
		\includegraphics[scale=0.3]{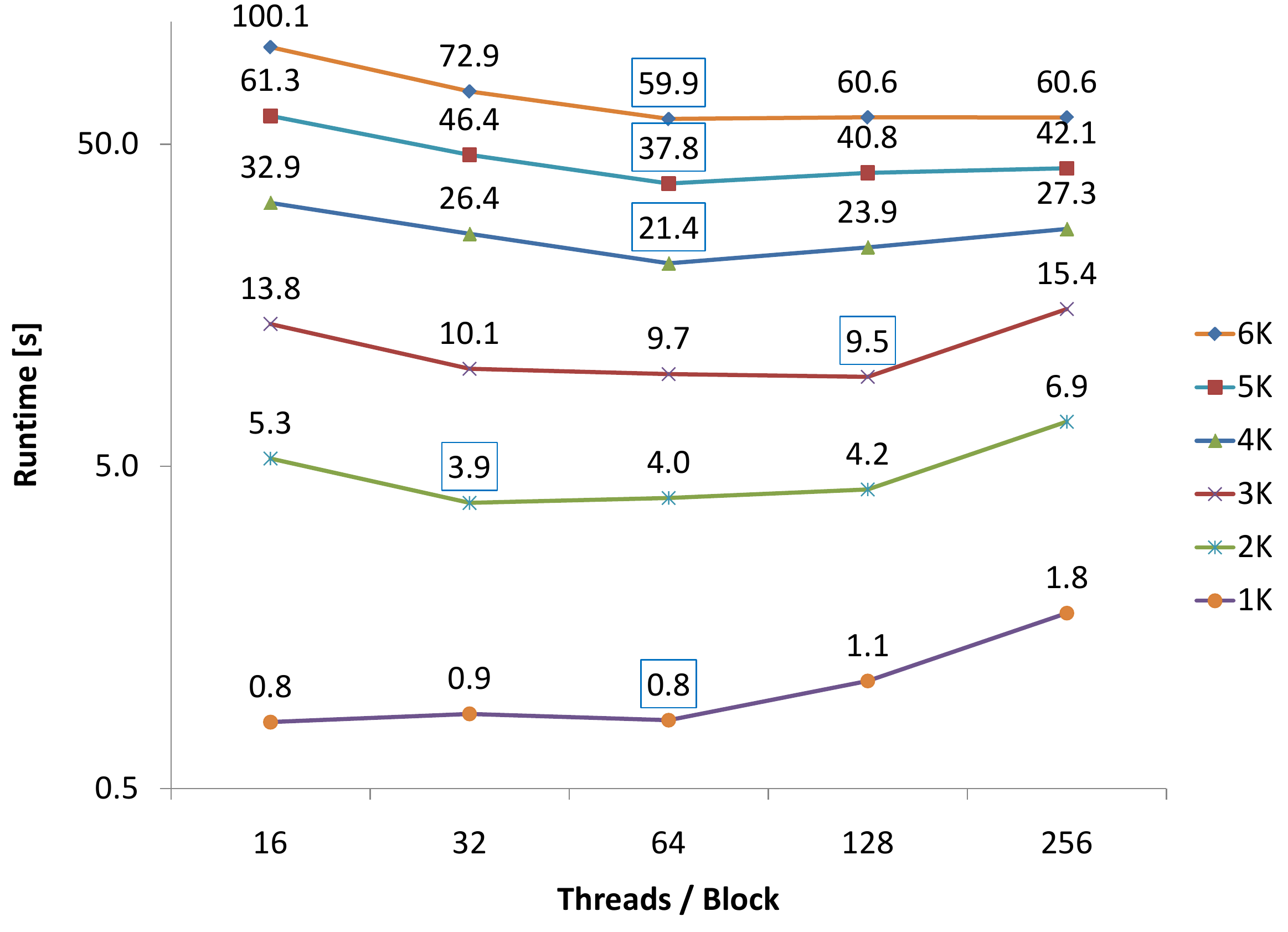}
		\label{fig:runtimes_gtx260}		
	}
  \hspace{0.2cm}
  \subfloat[GTX 480]
  {
		\includegraphics[scale=0.3]{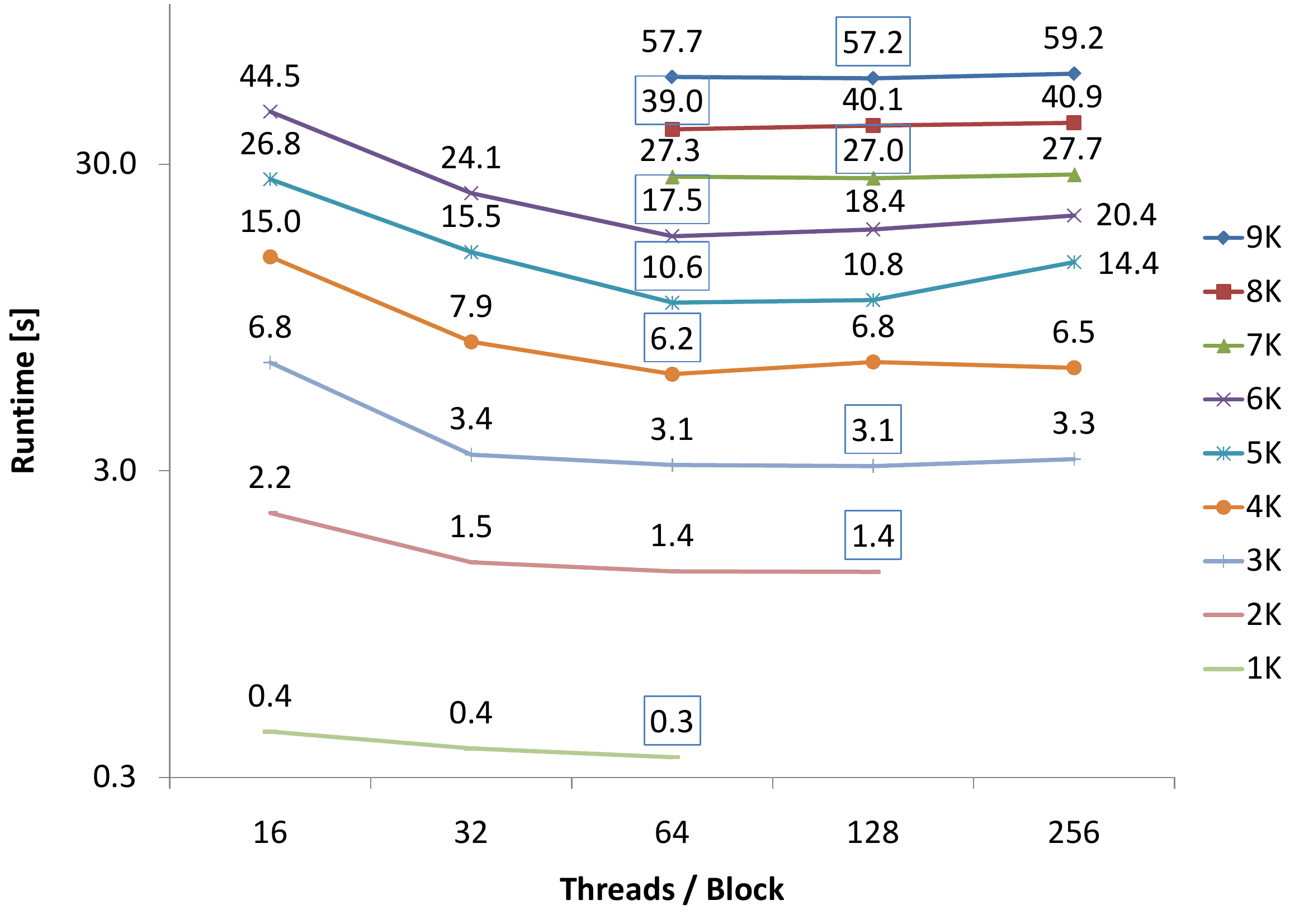}
		\label{fig:runtimes_gtx480}		
	}
  \caption{Runtimes for different thread counts per block. The input data size varies from 1K to 6K. The best runtime for each input size is displayed in a box.}
\end{figure}

Figures \ref{fig:runtimes_gtx260} and \ref{fig:runtimes_gtx480} show the runtimes (in seconds) for different numbers of threads in a block. The entire range of inputs is considered, including those that fit into memory only with output allocation disabled. The best (lowest) times are marked by a box. We observe that, for both cards, the best runtime is obtained generally with 64 threads per block. In the cases where this is not the case (usually for 128 threads), the difference is almost negligible.

The two chips powering the GTX 260 and GTX 480, have a related, but significantly different architecture. However, as observed from the figures, the configurations that best use their capabilities are similar, using 64 or 128 threads per block. Unlike usual CPU logic, running more threads than execution units is beneficial. This is due to the high latency in accessing the device global memory. By using many threads, the Block Scheduler can replace blocks waiting for memory fetching with those ready for execution. The result is a higher throughput due to high reuse of idle threads.

\subsection{Performance of \deltam  computation }

In this section we discuss the performance of the code on the two GPU platforms (from the latest two generations), with respect to the CPU implementation running on two different processors. The entire range of input sizes is tested with all variations of segment lengths. The best times are then selected and used for calculating the runtime improvement relative to the CPU implementation. Efficiency and GFLOPS for each graphics card are then computed.

The improvement factor from the GPU version is calculated against the reference Fortran MPI code running on the CPU. For AMD, the time duration obtained by running the program with 1 and 4 processes is used. The Intel i7-960 is equipped with Hyperthreading, meaning it can run 8 threads on just 4 physical cores. However, it was found that, in some cases, the 4 threads (MPI processes) version is faster. Therefore, one process and the best out of 4 and 8 processes is used as reference. The final runtime improvement factor for each input size is obtained by dividing the best time for each CPU by the best time of the GPU. When single-core is used as reference, the time measured while running the algorithm with just one process is divided by the best time of the GPU.
\begin{figure}[htb]
  \centering
  \subfloat[GTX 260]
  {
		\includegraphics[scale=0.3]{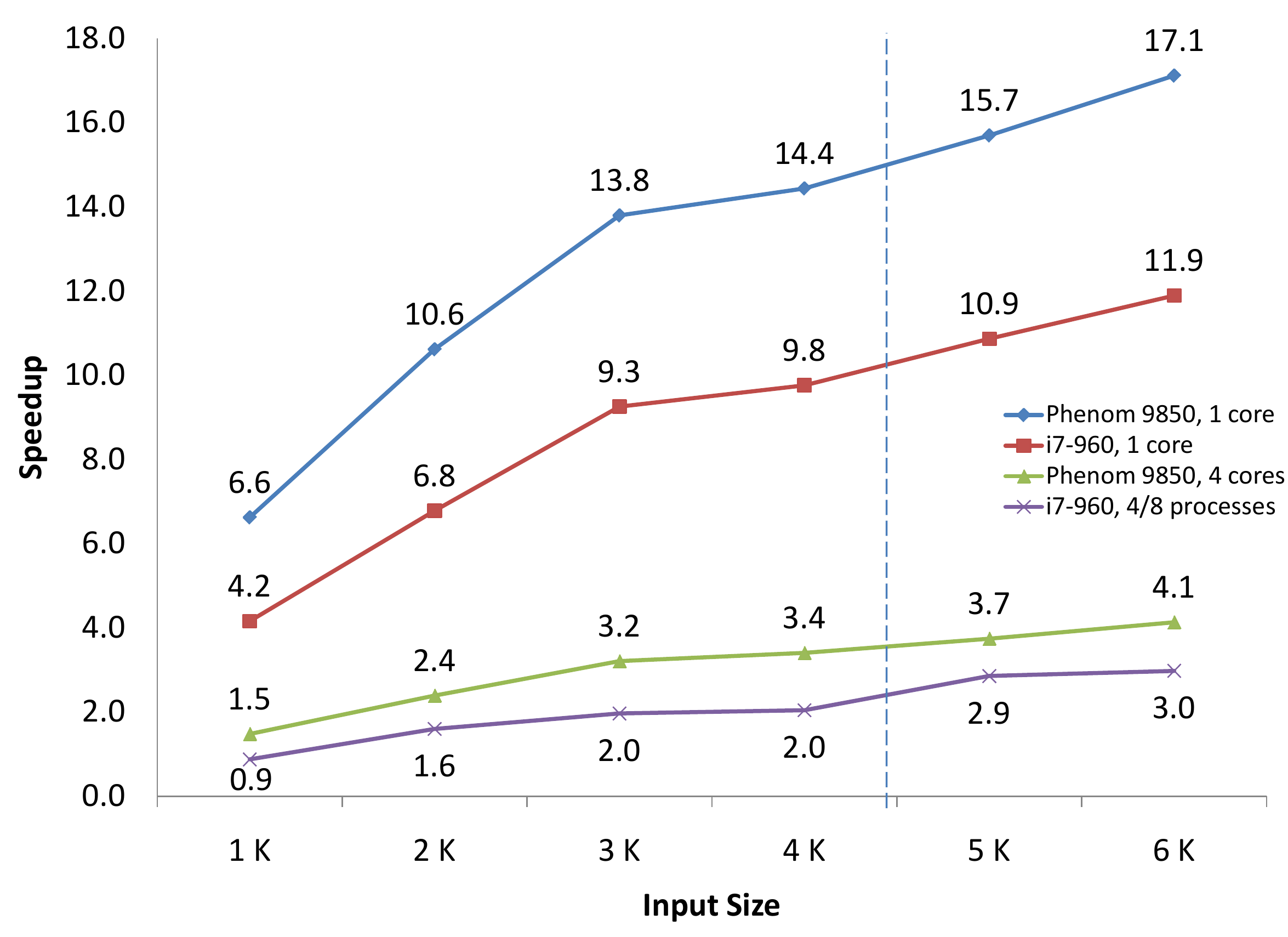}
		\label{fig:speedup_gtx260}		
	}
  \hspace{0.2cm}
  \subfloat[GTX 480]
  {
		\includegraphics[scale=0.3]{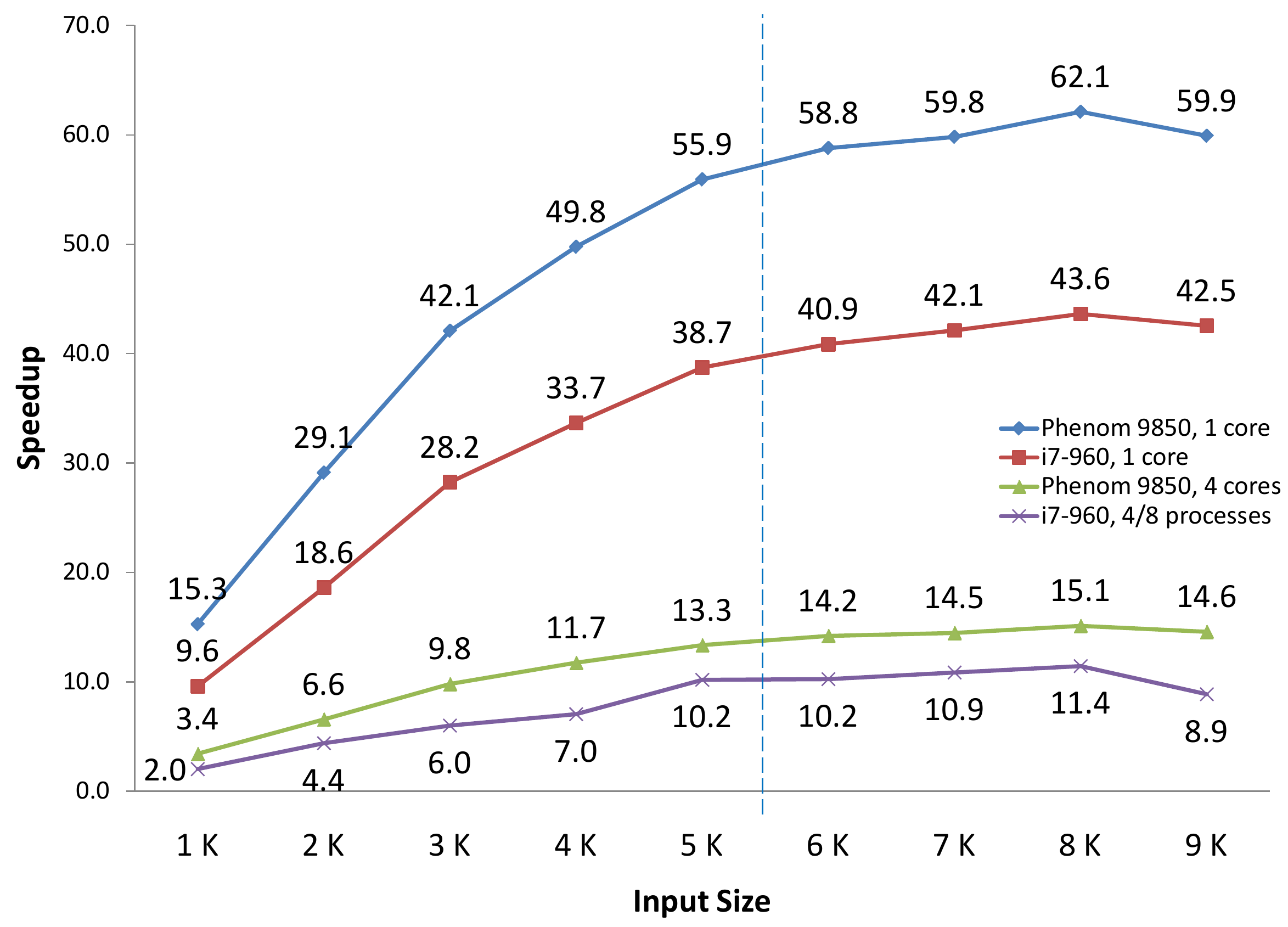}
		\label{fig:speedup_gtx480}		
	}
  \caption{Improvement factor obtained by \deltam{} calculation of \alm2map with CUDA with respect to the MPI version ran on the AMD Phenom and Intel i7 CPUs}
\end{figure}

Figures \ref{fig:speedup_gtx260} and \ref{fig:speedup_gtx480} show the runtime improvement for the two platforms used for testing (the latest generation GTX 480 and the older GTX 260) while using the entire range of inputs. We observe how larger inputs result in a higher improvement factor. Values rise sharply before starting to level at 4K (GTX 260) or 5K (GTX 480). The graphs plot the values for input sizes that normally fit the cards used for testing as well as those that require output disabling. They are separated by a vertical line (normal inputs on the left).

The AMD Phenom is slower than the Intel i7, therefore the improvement factor obtained will naturally be higher. When comparing the GTX 480 runtimes to those of single core CPU code, the performance ratio levels out at ~60x for the Phenom and at ~42x for the i7. For the older generation GTX 260, the factor is 3-3.5 times lower, at ~17x and  ~12x, respectively. However, the relevant values are those obtained when using the CPUs to their full potential, with all their cores. The algorithm scales almost perfectly with the number of physical cores, the improvement values being generally one fourth of single core, with ~14x and ~10x for GTX 480 and ~4x and ~3x for GTX 260. Intel Hyperthreading does not seem able to provide advantages by pushing scaling beyond the number of physical cores.

\begin{figure}
	\centering
		\includegraphics[scale=0.3]{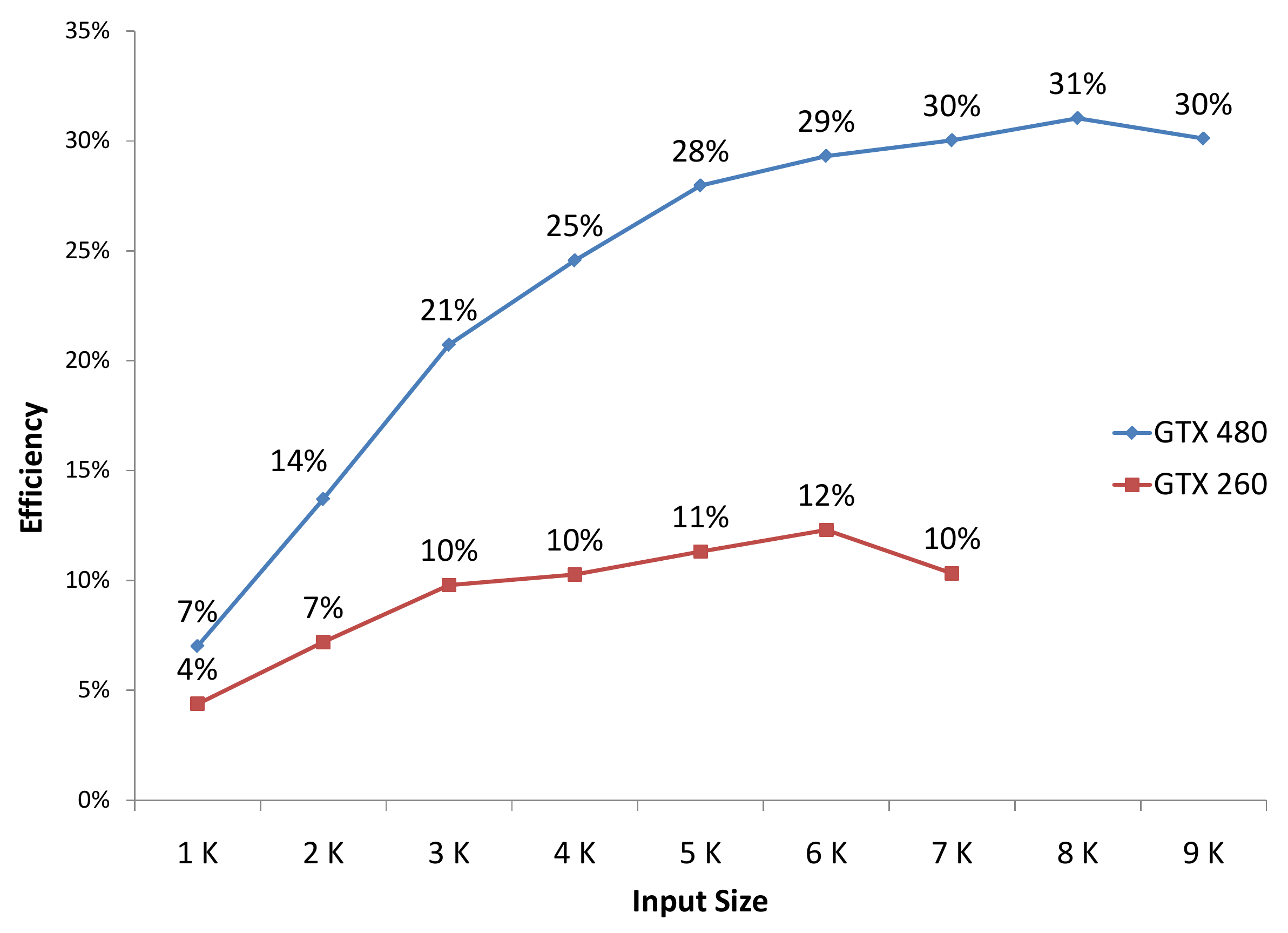}
	\caption{\alm2map efficiency on GPUs}
	\label{fig:efficiency}
\end{figure}

Figure \ref{fig:efficiency} displays the variation of efficiency with respect to the size of the input. The aspect of the curve is similar to that of the speedup graph, rising sharply before levelling at a 3K or 5K input. Being a memory-starved algorithm on the CPU, an adaptation to a faster chip, with a high-latency memory, was bound to suffer of the same problem. Running \s2hat on the Intel i7-960, it reaches an efficiency of just 10\%. As it can be seen in the graph, for the GTX 260, it mantains roughly the same value of 10\%. On the GTX 480, however, it improves by a factor of 3, peaking at 30\%. Since on the commercial GTX 480 double precision is limited to a quarter of its performance, running the algorithm on a Tesla card (which does not have it) will probably decrease efficiency by exposing the memory latency issue, hidden by this limitation, but should significantly improve performance.

\begin{figure}[htb]
  \centering
  \subfloat[GTX 260]
  {
		\includegraphics[scale=0.3]{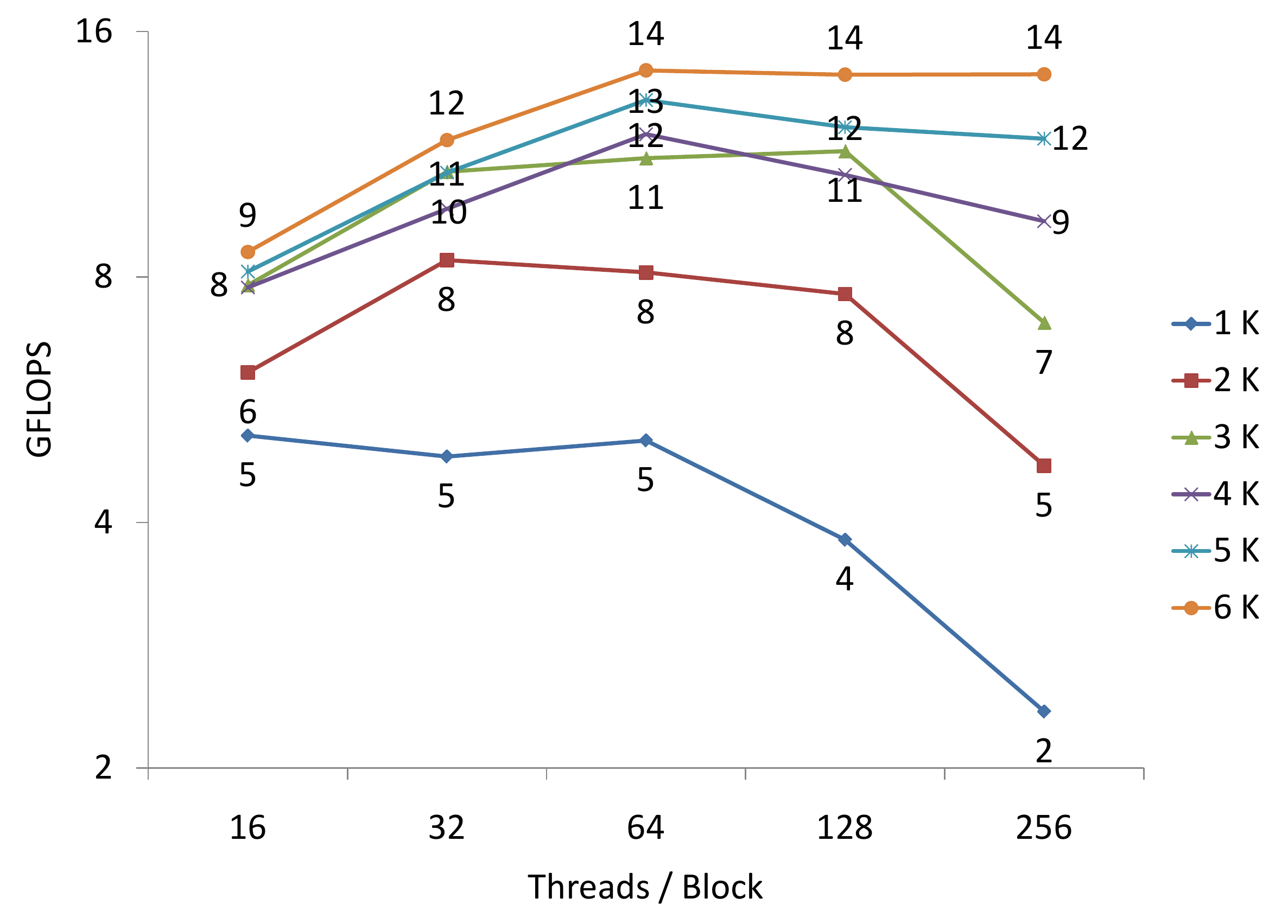}
		\label{fig:gflops_gtx260}		
	}
  \hspace{0.2cm}
  \subfloat[GTX 480]
  {
		\includegraphics[scale=0.3]{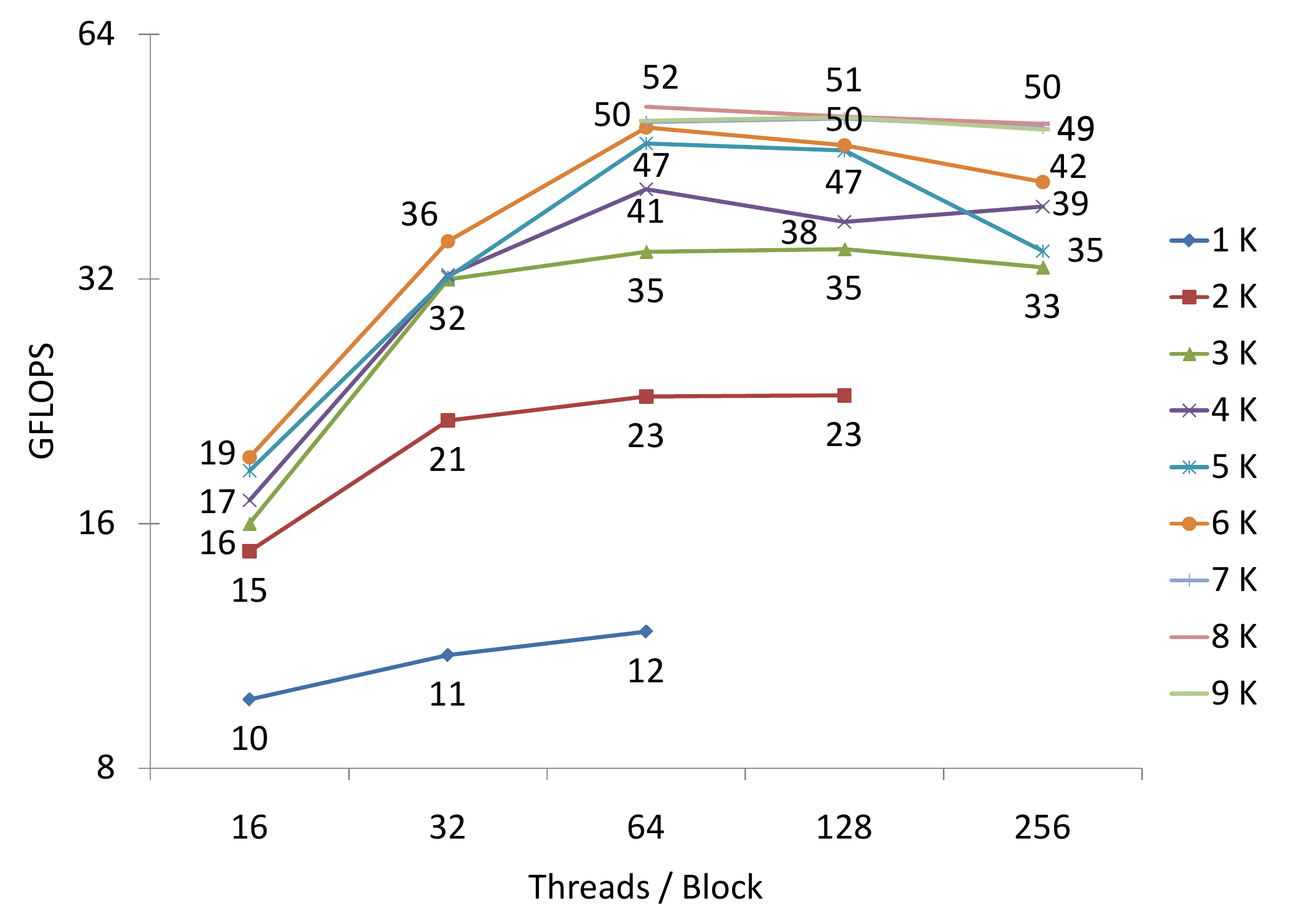}
		\label{fig:gflops_gtx480}		
	}
  \caption{GFLOPS per thread count}
\end{figure}

In figures \ref{fig:gflops_gtx260} and  \ref{fig:gflops_gtx480} we display the GFLOPS values obtained for each input size, for different counts of threads per block. Floating point operations methodology is as follows: additions and multiplications are computed as one operation each; due to lack of good references, divisions, square roots and logarithms are each counted as 20 operations. Rudimentary testing shows this value (20) to be an adequate estimate and, in either case, these operations combined are just 0.55\% of the total number of floating point operations counted. Since GPUs do not yet have hardware counters for floating point operations, the operations were counted manually (adding values to variables in each thread followed by summation for the final results).

As resulting from figures \ref{fig:gflops_gtx260} and  \ref{fig:gflops_gtx480}, the number of resulting GFLOPS increases with the size of the input, with values leveling at 50-52 GFLOPS for GTX 480 (figure \ref{fig:gflops_gtx480}) and 14 for GTX 260 (figure \ref{fig:gflops_gtx260})

\subsection{Overall performance}
The performance of the \alm2map algorithm is greatly improved by offloading the \deltam computation onto a GPU. However, the second step, FFT calculation, requires attention also. 
In the original CPU-only code, the FFTs occupy 5-10\% of the total runtime. Improving \deltam timing by a factor of 10 (Intel I7-960, 4 processes), results in the FFTs becoming dominant.

Since different FFT packages have different runtime characteristics, two CPU-only FFT libraries have been tested: one as implemented in Healpix~\cite{Gorski_etal_2005} and the other -- FFTW\footnote{FFTW: http://www.fftw.org/}~\cite{Frigo2005design}. Also, an FFT library for the NVIDIA GPUs, CUFFT \cite{CUFFT}, is employed. The current \alm2map FFT implementation is a direct port of the original CPU version, with no specific GPU optimizations.

\begin{figure}[htb]
  \centering
  \subfloat[]
  {
		\includegraphics[scale=0.3]{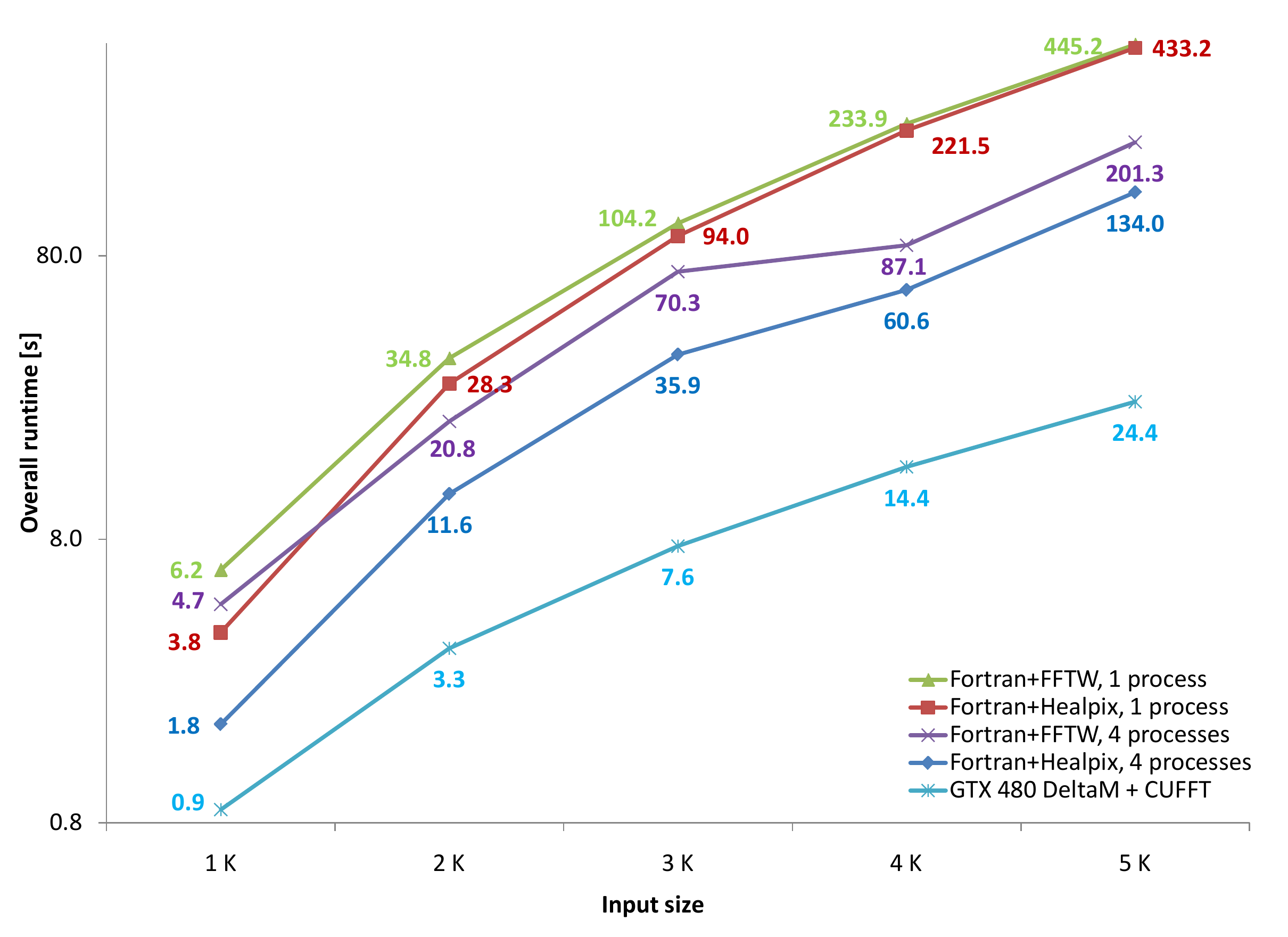}
		\label{fig:overall_runtime_a}		
	}
  \hspace{0.2cm}
  \subfloat[]
  {
		\includegraphics[scale=0.3]{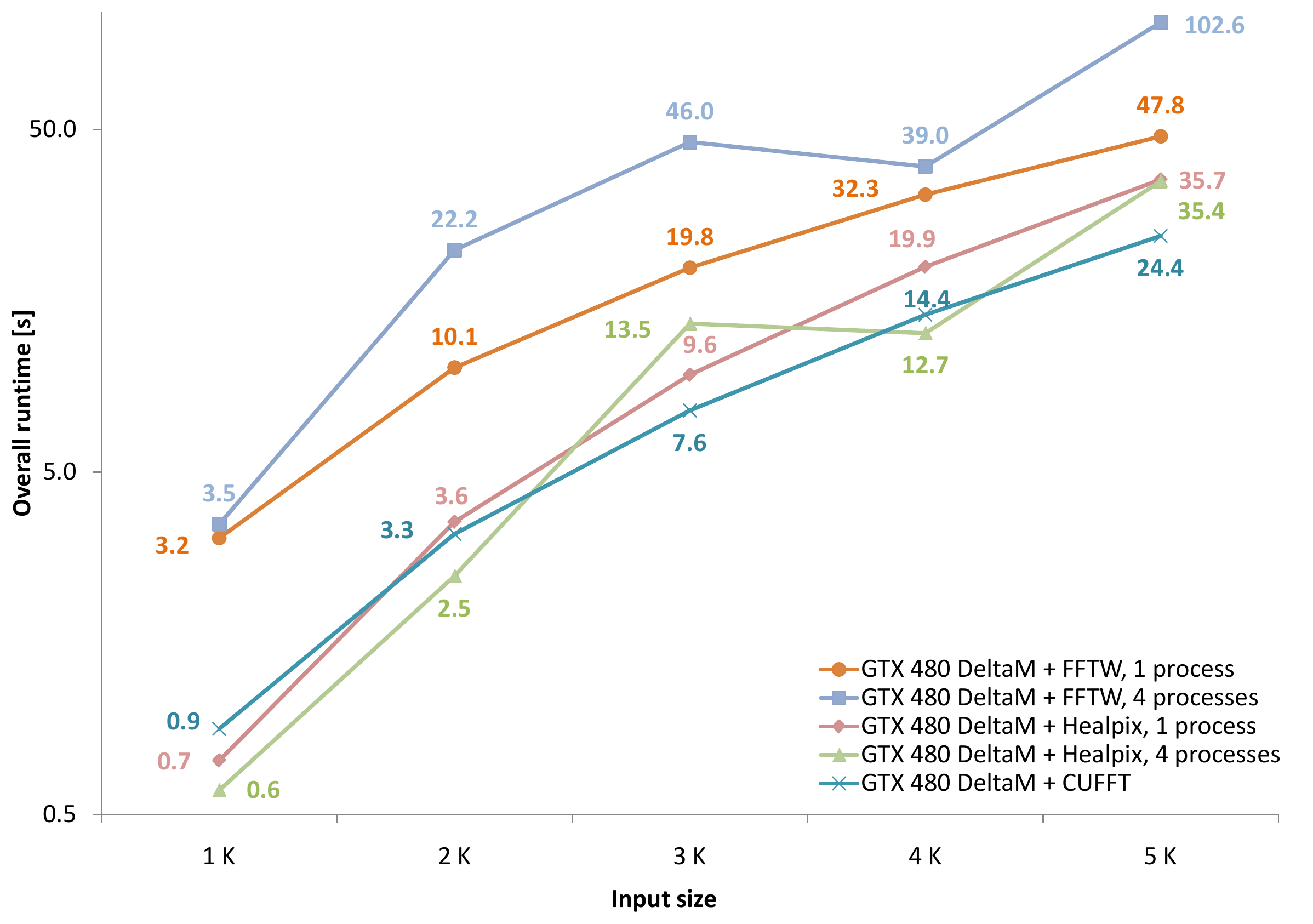}
		\label{fig:overall_runtime_b}		
	}
  \caption{\alm2map{} overall runtime, Intel i7-960 and NVIDIA GTX 480}
  \label{fig:overall_runtime}
\end{figure}

Figure \ref{fig:overall_runtime} shows the overall (\deltam+ FFT) runtimes for all combinations of \deltam computing code (Fortran on Intel i7-960 or CUDA on NVIDIA GTX 480), CPU FFT packages (Healpix or FFTW) and process count (1 or 4). Also, the runtime for a full GPU computation is plotted. Only the Intel i7-960 with NVIDIA GTX 480 results are shown.

We notice that, relative to the FFTW version, the Healpix package performs better for both 1 and 4 processes. We also observe that the best runtimes belong to the
code running on the GPU. 

\begin{figure}[htb]
  \centering
		\includegraphics[scale=0.33]{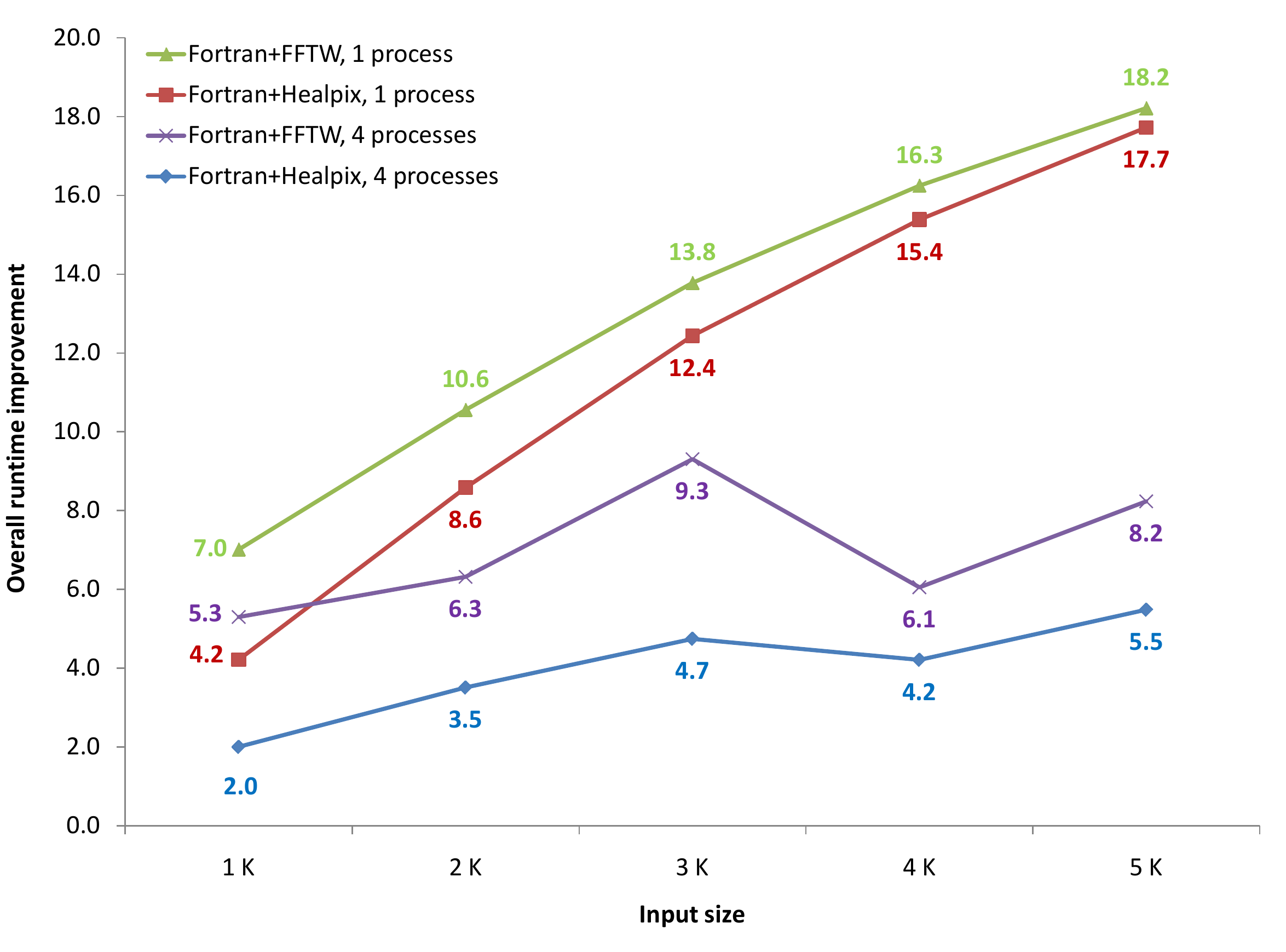}
		
  \caption{\alm2map{} overall performance improvement
\label{fig:overall_speedup}		
}
\end{figure}

Figure \ref{fig:overall_speedup} plots the overall runtime improvement over the CPU code versions with respect to the best performing GPU code (labeled ``GTX480 \deltam + CUFFT'' in figure   \ref{fig:overall_runtime}). We observe that, in the best case, the improvement is just half that obtained when considering just the \deltam computation (figure \ref{fig:speedup_gtx480}), but also significant, reaching factors from 5 to 18 (when comparing to the best and worst, respectively, performing code).

\section{Conclusions and Future Work}
\label{sec:concl}

This paper describes an algorithm for computing the inverse spherical
harmonic transform on GPUs.  The algorithm is compared with the
implementation of the inverse spherical harmonic transform provided in
\s2hat library, implemented using Fortran and MPI.  The GPU algorithm
leads to an improvement of up to a factor of $18$ with respect to
\s2hat on a single core and up to a factor of $5.5$ with respect to
\s2hat on 4 cores of an Intel i7-960 machine.  The improvement is
limited by the performance of Fast Fourier transforms.


Even though a single GPU offers high computing power, employing
several is the easiest way of scaling performance as well as handling
larger inputs. The algorithm has been designed with multi-GPU use in
mind and it can directly fit into, and benefit from, the \s2hat MPI
structure enabling straightforwardly distributed GPU computing.
However special care has to be taken to ensure a good load balance
among processors. This is the object of our current work.



\section*{Acknowledgments}
This work has been supported in part by French National Research Agency (ANR)
through COSINUS program (project MIDAS no. ANR-09-COSI-009).

\nocite{*} 
\bibliographystyle{plain} {
 \bibliography{sections/gpus} 

\begin{thebibliography}{10}

\bibitem{ArfkenBook}
G.~B. {Arfken} and H.~J. {Weber}.
\newblock {\em {Mathematical methods for physicists 6th ed.}}
\newblock {Academic Press}, 2005.

\bibitem{Bennett_etal_2003}
C.~L. {Bennett}, M.~{Bay}, M.~{Halpern}, G.~{Hinshaw}, C.~{Jackson},
  N.~{Jarosik}, A.~{Kogut}, M.~{Limon}, S.~S. {Meyer}, L.~{Page}, D.~N.
  {Spergel}, G.~S. {Tucker}, D.~T. {Wilkinson}, E.~{Wollack}, and E.~L.
  {Wright}.
\newblock {The Microwave Anisotropy Probe Mission}.
\newblock {\em \apj}, 583:1--23, January 2003.

\bibitem{Doroshkevich_etal_2005}
A.~G. {Doroshkevich}, P.~D. {Naselsky}, O.~V. {Verkhodanov}, D.~I. {Novikov},
  V.~I. {Turchaninov}, I.~D. {Novikov}, P.~R. {Christensen}, and {L.~-.}
  {Chiang}.
\newblock {First Release of Gauss-Legendre Sky Pixelization (GLESP) software
  package for CMB analysis}.
\newblock {\em ArXiv Astrophysics e-prints}, January 2005.

\bibitem{Driscoll_Healy_1994}
J.~R. Driscoll and D.~M. Healy.
\newblock Computing fourier transforms and convolutions on the 2-sphere.
\newblock {\em Advances in Applied Mathematics}, 15(2):202 -- 250, 1994.

\bibitem{Frigo2005design}
M.~Frigo and S.G. Johnson.
\newblock {The design and implementation of FFTW3}.
\newblock {\em Proceedings of the IEEE}, 93(2):216--231, 2005.

\bibitem{Gorski_etal_2005}
K.~M. {G{\'o}rski}, E.~{Hivon}, A.~J. {Banday}, B.~D. {Wandelt}, F.~K.
  {Hansen}, M.~{Reinecke}, and M.~{Bartelmann}.
\newblock {HEALPix: A Framework for High-Resolution Discretization and Fast
  Analysis of Data Distributed on the Sphere}.
\newblock {\em \apj}, 622:759--771, April 2005.

\bibitem{Muciaccia_etal_1997}
P.~F. {Muciaccia}, P.~{Natoli}, and N.~{Vittorio}.
\newblock {Fast Spherical Harmonic Analysis: A Quick Algorithm for Generating
  and/or Inverting Full-Sky, High-Resolution Cosmic Microwave Background
  Anisotropy Maps}.
\newblock {\em \apjl}, 488:L63+, October 1997.

\bibitem{Nukada09}
Akira Nukada and Satoshi Matsuoka.
\newblock Auto-tuning 3-{D} {FFT} library for {CUDA} {GPU}s.
\newblock In {\em SC '09: Proceedings of the Conference on High Performance
  Computing Networking, Storage and Analysis}, pages 1--10, New York, NY, USA,
  2009. ACM.

\bibitem{CUFFT}
Nvidia.
\newblock {\em CUDA CUFFT Library}, 2010.

\bibitem{CUDABestPractices}
Nvidia.
\newblock {\em NVIDIA CUDA Best Practices Guide}, 2010.

\bibitem{CUDAManual}
Nvidia.
\newblock {\em NVIDIA CUDA Programming Guide}, 2010.

\bibitem{CUDAFermi}
Nvidia.
\newblock {\em Tuning CUDA Applications for Fermi}, 2010.

\bibitem{Smoot_etal_1992}
G.~F. {Smoot}, C.~L. {Bennett}, A.~{Kogut}, E.~L. {Wright}, J.~{Aymon}, N.~W.
  {Boggess}, E.~S. {Cheng}, G.~{de Amici}, S.~{Gulkis}, M.~G. {Hauser},
  G.~{Hinshaw}, P.~D. {Jackson}, M.~{Janssen}, E.~{Kaita}, T.~{Kelsall},
  P.~{Keegstra}, C.~{Lineweaver}, K.~{Loewenstein}, P.~{Lubin}, J.~{Mather},
  S.~S. {Meyer}, S.~H. {Moseley}, T.~{Murdock}, L.~{Rokke}, R.~F. {Silverberg},
  L.~{Tenorio}, R.~{Weiss}, and D.~T. {Wilkinson}.
\newblock {Structure in the COBE differential microwave radiometer first-year
  maps}.
\newblock {\em \apjl}, 396:L1--L5, September 1992.

\bibitem{Tygert2008}
Mark Tygert.
\newblock Fast algorithms for spherical harmonic expansions, ii.
\newblock {\em Journal of Computational Physics}, 227(8):4260 -- 4279, 2008.

\bibitem{volkov08:_bench_gpus_to_tune_dense_linear_algeb}
V.~Volkov and J.~W. Demmel.
\newblock Benchmarking {GPU}s to tune dense linear algebra.
\newblock {\em ACM/IEEE Conference on Supercomputing (SC08)}, 2008.

\bibitem{Volkov08}
V.~Volkov and J.W. Demmel.
\newblock {LU}, {QR} and {C}holesky factorizations using vector capabilities of
  {GPU}s.
\newblock Technical Report UCB/EECS-2008-49, EECS Department, University of
  California, Berkeley, May 2008.

\end{thebibliography}
}

\end{document}